\documentclass[pdflatex,sn-mathphys-num]{sn-jnl}%

\usepackage{graphicx}%
\usepackage{multirow}%
\usepackage{amsmath,amssymb,amsfonts}%
\usepackage{amsthm}%
\usepackage{mathrsfs}%
\usepackage[title]{appendix}%
\usepackage{xcolor}%
\usepackage{textcomp}%
\usepackage{manyfoot}%
\usepackage{booktabs}%
\usepackage{algorithm}%
\usepackage{algorithmicx}%
\usepackage{algpseudocode}%
\usepackage{listings}%
\begin{document}

	\title[Self-Bound Anisotropic Stars in $\mathcal{R}+\alpha\mathcal{R}^2$ Gravity with a Hayward Exterior]{\texorpdfstring{ Self-Bound Anisotropic Stars in $\mathcal{R}+\alpha\mathcal{R}^2$ Gravity with a Hayward Exterior: Junction Conditions and NICER Constraints}{Self-Bound Anisotropic Stars in R + alpha R2 Gravity with a Hayward Exterior: Junction Conditions and NICER Constraints}}

	\author[1]{\fnm{Furqan} \sur{Nazeer}}\email{fnazeer.phdmath21sns@student.nust.edu.pk}

	\author*[1]{\fnm{Tooba} \sur{Feroze}}\email{tferoze@sns.nust.edu.pk}

	\affil[1]{\orgdiv{School of Natural Sciences}, \orgname{National University of Sciences and Technology}, \orgaddress{\street{H-12}, \city{Islamabad}, \country{Pakistan}}}

	\abstract{\noindent Analytic compact-star models in modified gravity usually borrow the junction conditions of general relativity and fix their constants with radii derived for a particular stellar model. Here we avoid both. We study $\mathcal{F}(\mathcal{R})=\mathcal{R}+\alpha\mathcal{R}^{2}$ gravity with a Durgapal--Bannerji interior of embedding class one (Karmarkar condition) and a Hayward exterior. Without Birkhoff's theorem, the exterior is an assumption, and we measure its effect with the Hayward metric, which deforms Schwarzschild by a length $\ell$. The junction conditions of quadratic $\mathcal{F}(\mathcal{R})$ gravity fix all constants in closed form in terms of $M$, $R$ and $\ell$. The condition $p_r(R)=0$ is replaced by continuity of the trace of the extrinsic curvature. The Ricci scalar jumps at the surface, so there is a thin shell with negative surface energy density for $\alpha>0$, a gravitational double layer, and a small radial stress. For this stress to be below 1\% of the central pressure we need $\alpha\lesssim0.2$~km$^{2}$ and $\ell\lesssim0.3M$. The interior behaves as a self-bound star. A single surface energy density, $\rho_s=(1.37\pm0.16)\,\rho_0$, reproduces PSR~J0740+6620, J0437$-$4715, J0030+0451 and J0614$-$3329, and for $\alpha=0.1$~km$^{2}$ the model predicts $R_{1.4}=11.5$~km and a maximum mass of $2.9\,M_\odot$. Because $\ell$ enters surface observables only at order $(\ell/M)^{2}(M/R)^{3}$, the results carry over to a Schwarzschild exterior. Calibrating instead with strange-star radii of eclipsing X-ray pulsars gives $R_{1.4}\approx9$~km, which current data rule out.}


	\maketitle
	\section{Introduction}\label{1s}
	Modern cosmological observations imply that the Universe is undergoing accelerated expansion. This is supported by supernova surveys, by the angular power spectrum of the cosmic microwave background and by baryon acoustic oscillation data \citep{c3,c2,c1,c5}. In the standard cosmological model the acceleration is caused by dark energy, a component with large negative pressure which dominates the present Universe. According to the Planck results, about $68.5\%$ of the present energy budget is in dark energy, about $4.9\%$ in baryonic matter and the remainder in dark matter \citep{Planck2018}, and these numbers are now well measured. In general relativity (GR), with the Einstein--Hilbert action, dark energy has to be a cosmological constant $\Lambda$, but the observed $\Lambda$ is many orders of magnitude below the value suggested by theory. This discrepancy is one of the main motivations for studying modified gravitational actions.

	\noindent The $\mathcal{F}(\mathcal{R})$ gravity model generalizes the Einstein–Hilbert action by introducing a function of the Ricci scalar, $\mathcal{R}$ \citep{c6,SotiriouFaraoni2010,DeFeliceTsujikawa2010}. \citet{c7} introduced another important modification, the $\mathcal{F}(\mathcal{G})$ theory, which is formulated using the Gauss-Bonnet invariant $\mathcal{G}$. Later, $\mathcal{F}(\mathcal{R},\mathcal{G})$ gravity was proposed, in which the action depends on both $\mathcal{R}$ and $\mathcal{G}$ \citep{c8}. Among the various modified gravity models, $\mathcal{F}(\mathcal{R},\mathcal{T})$ gravity couples the Ricci scalar $\mathcal{R}$ explicitly to the trace $\mathcal{T}$ of the energy--momentum tensor \citep{c9}. This matter--geometry coupling widens the range of problems the theory can address. A dust fluid in $\mathcal{F}(\mathcal{R},\mathcal{T})$ cosmology, for example, can give the $\Lambda$CDM expansion as well as phantom and phantom-to-non-phantom phases \citep{Jamil2012}. In astrophysics the coupling makes test particles move on non-geodesic paths, and in dense stars it changes the pressure balance by an amount that depends on the local matter content. Effects of this type have been studied in many modified theories and for many kinds of stars \citep{h4, c17, c16, c15,c12,c10}, and the same theories have also been used for strong-field objects other than stars. For instance, exact static wormhole solutions of $\mathcal{F}(\mathcal{R})$ gravity have been found for noncommutative-geometry sources, in which a point-like source is replaced by a smeared matter distribution \citep{Jamil2014}.\\
	\noindent Compact objects are well suited for testing gravity in the strong-field regime. They are formed at the end of stellar evolution, when a sufficiently massive star has used up its nuclear fuel and neither thermal pressure nor electron degeneracy pressure can stop the gravitational collapse \citep{LattimerPrakash2004}. A neutron star has a mass of one to two solar masses and a radius of about ten kilometers. Such extreme conditions cannot be reached in terrestrial experiments, and compact stars therefore also provide a test of the modified gravity theories proposed to explain the accelerated expansion. It is, however, difficult to obtain interior solutions which are analytically tractable and physically acceptable at the same time. \citet{c49} examined 127 candidate solutions in GR and found that only 16 of them pass all elementary tests of physical acceptability, and only 9 of these also have a sound speed that decreases monotonically outward. A useful approach is to choose a physically motivated ansatz for one metric function and to determine the other one from a geometric condition. The Durgapal--Bannerji ansatz \citep{dur} is a reliable choice. It is free from singularities and has been used successfully in several modified theories of gravity \citep{mau,dit,ven}. A four-dimensional spacetime can be embedded in a five-dimensional pseudo-Euclidean space if its metric potentials satisfy the Karmarkar condition \citep{c34}, making it a Class-I manifold. The two metric functions are related through an ordinary differential equation. Hence no equation of state needs to be imposed, and models of this type have already been found to be physically consistent in $\mathcal{F}(\mathcal{R})$ gravity \citep{ahm, nazt, naz}.

    \noindent Most analytic models of this kind are based on three simplifications (for a review of stellar models in modified gravity, see \citet{Olmo2020}). (i) The exterior is taken to be the Schwarzschild metric or, in more recent papers, a regular black-hole metric such as the Bardeen metric \citep{c42,h12,h13,h14,h15,c48}, although Birkhoff's theorem is not valid in $\mathcal{F}(\mathcal{R})$ gravity and the theory does not determine the exterior. (ii) Interior and exterior are matched with the junction conditions of GR, usually by setting $p_r(R)=0$, while quadratic $\mathcal{F}(\mathcal{R})$ gravity requires more conditions \citep{Deruelle2008,Senovilla2013}. (iii) The constants of the model are fixed with radii of X-ray pulsars that were themselves derived with a strange-star equation of state \citep{Gangopadhyay2013}. This last step is not needed anymore, because the Neutron Star Interior Composition Explorer (NICER) now provides radii of millisecond pulsars from pulse-profile modeling, without any assumption on the equation of state \citep{Salmi2024J0740,Choudhury2024}.

\noindent In this paper we address these three points for the Durgapal--Bannerji--Karmarkar interior in $\mathcal{F}(\mathcal{R})=\mathcal{R}+\alpha\mathcal{R}^{2}$ gravity with a Hayward exterior. We use the junction conditions of quadratic $\mathcal{F}(\mathcal{R})$ gravity \citep{Senovilla2013}. The continuity of the trace of the extrinsic curvature fixes the integration constant $b$ in closed form and replaces the condition $p_r(R)=0$. However, the Ricci scalar still has a jump at the surface. This jump gives a thin shell and a gravitational double layer with negative surface energy density for $\alpha>0$, and it leaves a radial stress at the surface which limits both $\alpha$ and $\ell$. The model is then tested against NICER and radio-timing data. One surface energy density, $\rho_s=(1.37\pm0.16)\,\rho_0$, reproduces PSR~J0740+6620, J0437$-$4715, J0030+0451 and J0614$-$3329 together, with $R_{1.4}=11.5$~km and a maximum mass of $2.9\,M_\odot$. The radius of the model increases with mass between 1 and $2\,M_\odot$, which can be tested by NICER observations of light pulsars. We also consider the Hayward metric as a controlled, asymptotically flat, one-parameter deformation of the Schwarzschild metric and show analytically that its effect on surface observables is of order $\beta^{2}(M/R)^{3}$, with $\beta=\ell/M$. Therefore the comparison with observations is also valid for a Schwarzschild exterior, and $\ell$ is bounded by the junction conditions.
\\The paper is organized as follows. In Section~\ref{2s} we give the field equations of $\mathcal{F}(\mathcal{R})$ gravity for a static, spherically symmetric anisotropic star. Section~\ref{3s} contains the class-one interior (Sec.~\ref{sec:interior}), the Hayward exterior (Sec.~\ref{sec:exterior}) and the junction conditions (Sec.~\ref{sec:junction}). In Section~\ref{4s} the physical admissibility of the interior is checked for the observed pulsars, and in Section~\ref{sec:obs} we derive the allowed values of $\alpha$ and $\ell$ and compare the model with measured masses and radii. The results and their limitations are discussed in Section~\ref{sec:disc}, and the conclusions are given in Section~\ref{5s}.

\section{Field Equations in the $\mathcal{F}(\mathcal{R})$ Model}\label{2s}

The foundation of GR lies in the Einstein--Hilbert action, where the spacetime dynamics are governed by the Ricci scalar $\mathcal{R}$. This action is expressed as
\begin{equation}\label{e1}
	S=\frac{1}{2k}\int \mathcal{R}\sqrt{-g}\, d^4x,
\end{equation}
with $g$ denoting the determinant of the metric tensor $g_{ij}$ and $k=\frac{8\pi G}{c^4}$.
This action can be generalized by replacing $\mathcal{R}$ with a general function $\mathcal{F}(\mathcal{R})$. In this setting, the action becomes

\begin{equation}\label{e2}
	S=\frac{1}{2k}\int\mathcal{F}(\mathcal{R})\sqrt{-g}\,d^{4}x+\int\mathcal{L}_m\sqrt{-g}\,d^{4}x ,
\end{equation}
where $\mathcal{L}_m$ encodes the matter content.\\
Carrying out the variation of the total action \eqref{e2} yields the modified field equations
\begin{equation}\label{6e}
	(R_{ij}-\nabla_i\nabla_j+g_{ij}\Box)\mathcal{F}_{\mathcal{R}}
	-\frac{1}{2}g_{ij}\mathcal{F}
	=8\pi T_{ij},
\end{equation}
where $T_{ij}$ is the energy momentum tensor  and $\mathcal{F}_{\mathcal{R}}=\partial \mathcal{F}/\partial \mathcal{R}$. The operator $\Box$ represents the d'Alembertian,
\begin{equation}
	\Box=\frac{1}{\sqrt{-g}}\frac{\partial}{\partial x^i}\left(\sqrt{-g}g^{ij}\frac{\partial}{\partial x^j}\right).
\end{equation}
The interior geometry of the compact object is described by the static, spherically symmetric line element
\begin{equation}\label{13e}
	ds^2=-A^2(r)\,dt^2+B^2(r)\,dr^2+r^2(d\theta^2+\sin^2\theta\,d\phi^2).
\end{equation}
The pressure inside compact stars can be anisotropic for several reasons, for example strong interactions, phase transitions and magnetic effects \citep{h10}. A difference between the radial and tangential pressures has an important influence on the equilibrium and stability of the star, and therefore we describe the stellar interior by an anisotropic fluid. The corresponding energy momentum tensor is written as
\begin{equation}
	T_{ij}=(\rho+p_t)u_i u_j + p_t g_{ij} + (p_r-p_t)v_i v_j,
\end{equation}
where $\rho$, $p_r$, and $p_t$ denote the energy density, radial pressure, and tangential pressure, respectively. The four-velocity satisfies $u_i u^i=-1$, and $v_i$ is a spacelike unit vector.
\\For practical applications, we restrict attention to the Starobinsky model \citep{SM}
\begin{equation}
	\mathcal{F}(\mathcal{R})=\mathcal{R}+\alpha \mathcal{R}^2,
\end{equation}
here $\alpha$ is a constant. The standard Einstein theory is recovered by taking $\alpha = 0$. We use units with $G=c=1$, lengths in km, $\alpha$ in km$^{2}$ and $GM_\odot/c^{2}=1.4766$~km. The scalaron is stable only for $\mathcal{F}''=2\alpha>0$. Laboratory experiments give the bound $\alpha\lesssim10^{-10}$~m$^{2}$, and Gravity Probe~B and the double pulsar give the much weaker bounds $\alpha\lesssim5\times10^{11}$~m$^{2}$ and $2.3\times10^{15}$~m$^{2}$, respectively \citep{NafJetzer2010}. A universal $\alpha$ of order km$^{2}$, that is $10^{6}$~m$^{2}$, is therefore not allowed. Earlier compact-star studies in this theory have nevertheless worked with values of this size \citep{Yazadjiev2014}, and we follow them, regarding $\alpha$ as an effective coupling in the strong-field regime that is to be determined from the stars. With this choice, the field equations \eqref{6e} for the anisotropic configuration in  static, spherically symmetric space time become
\begin{equation}\label{15e}
	\begin{split}
		8\pi\rho&=(1+2\alpha \mathcal{R})\left[\frac{1}{B^2}\left( \frac{2B'}{rB}-\frac{1}{r^2}\right)+\frac{1}{r^2} \right]+\frac{2\alpha}{B^2}\left[ \mathcal{R}'\left(\frac{B'}{B}-\frac{2}{r}\right)-\mathcal{R}''\right]-\frac{\alpha \mathcal{R}^2}{2},\\
		8\pi p_r&=(1+2\alpha \mathcal{R})\left[\frac{1}{B^2}\left( \frac{2A'}{rA}+\frac{1}{r^2}\right)-\frac{1}{r^2} \right]+\frac{2\alpha}{B^2}\left[ \mathcal{R}'\left(\frac{A'}{A}+\frac{2}{r}\right) \right] +\frac{\alpha \mathcal{R}^2}{2},\\
		8\pi p_t&=(1+2\alpha \mathcal{R})\left[\frac{1}{B^2}\left(\frac{A''}{A}-\frac{A'B'}{AB}+\frac{A'}{rA}-\frac{B'}{rB}\right) \right] +\frac{2\alpha}{B^2}\left[\mathcal{R}'\left(\frac{A'}{A}-\frac{B'}{B}+\frac{1}{r}\right)+\mathcal{R}''\right]+\frac{\alpha \mathcal{R}^2}{2},
	\end{split}
\end{equation}
and
\begin{equation}
	\mathcal{R}=\frac{2}{B^2}\left(\frac{A'B'}{AB}+\frac{2B'}{rB}-\frac{A''}{A}-\frac{2A'}{rA}-\frac{1}{r^2}+\frac{B^2}{r^2}\right),
\end{equation}
where the primes ($\prime$) represent differentiation with respect to the radial coordinate $r$.

\section{\texorpdfstring{Hayward Matched Class-I Model in $\mathcal{F}(\mathcal{R})$}{Hayward Matched Class-I Model in F(R)}}\label{3s}

\subsection{Class-one interior}\label{sec:interior}

Obtaining exact solutions in $\mathcal{F}(\mathcal{R})$ gravity is rarely straightforward. For compact stars, one successful route is to impose the Karmarkar condition and to choose a suitable form for one of the metric potentials. The field equations (\ref{15e}) contain the five unknowns $A(r)$, $B(r)$, $\rho$, $p_r$ and $p_t$, so that additional assumptions are needed. Normally either two of the unknowns are fixed directly or a relation between them is assumed.
\\In this study, we develop a new anisotropic compact-star configuration in the context of modified gravity theory while avoiding unnecessary assumptions. As a starting point, the Karmarkar condition is imposed to establish a relationship between the metric potentials. A spacetime that satisfies this condition is classified as an embedding Class-I geometry, meaning that the corresponding four-dimensional manifold can be embedded in a five-dimensional pseudo-Euclidean space \citep{c34}. The condition is expressed as
\begin{equation}\label{17e}
	R_{2323}R_{1010}=R_{1212}R_{3030}+R_{1220}R_{1330}.
\end{equation}
Using the non-zero curvature terms appearing in Eq.~(\ref{17e}), one can derive the following relationship between the metric functions \citep{new}
\begin{equation}\label{18e}
	A(r)=a+b\int \sqrt{B^2(r)-1}\, dr,
\end{equation}
here $a$ and $b$ are constants of integration.
With the aim of generating a viable Class-I stellar model, the following metric function is introduced as \citep{dur}
\begin{equation}\label{19e}
	B(r)=\sqrt{\frac{2(1+cr^2)}{2-cr^2}},
\end{equation}
with $c>0$. This form is chosen since it gives a realistic stellar behavior and is compatible with the modified gravity theory used here. The function $B(r)$ increases smoothly with $r$ and is regular at the center, where $B(r)|_{r=0}=1$. Substituting this metric function in Eq.~(\ref{18e}) gives
\begin{equation}
	A(r)=a-b\sqrt{\frac{6-3cr^2}{c}}
\end{equation}
This metric potential also behaves regularly throughout the stellar interior and increases monotonically. Regular metric potentials are needed here, since $\rho$, $p_r$ and $p_t$ are calculated from $A(r)$, $B(r)$ and their derivatives and would otherwise behave unphysically.\\
\noindent Using the above ansatz in Eq.~(\ref{15e}), the expressions for the density, radial pressure, and tangential pressure take the form

\begin{equation}\label{21e}
	\begin{split}
		\rho&=\frac{1}{16\pi}\left[2f_3(r)\cdot(1+2\alpha h_1(r))+4\alpha f_1(r)\cdot(h_2(r)g_1(r)-h_3(r))-\alpha (h_1(r))^2\right],\\
		p_r&=\frac{1}{16\pi}\left[2f_4(r)\cdot(1+2\alpha h_1(r))+4\alpha f_1(r)h_2(r)g_2(r)+\alpha (h_1(r))^2\right],\\
		p_t&=\frac{1}{16\pi}\left[2f_5(r)\cdot(1+2\alpha h_1(r))+4\alpha f_1(r)\cdot(h_2(r)g_3(r)+h_3(r))+\alpha (h_1(r))^2\right],\\
	\end{split}
\end{equation}
where
\begin{equation}
	\begin{split}
		f_3(r)&=\frac{3c\left(3+cr^{2}\right)}{2\left(1+cr^{2}\right)^{2}},
		\hspace{0.5cm}
		f_4(r)=\frac{c\left(5b f_2(r)-3a\right)}{2\left(1+cr^{2}\right)\left(a-b f_2(r)\right)}, \hspace{0.5cm} 	f_5(r)=\frac{c\left[b\left(5+cr^{2}\right)f_2(r)-3a\right]}{2\left(1+cr^{2}\right)^{2}\left(a-bf_2(r)\right)},
		\\[0.3cm]
		g_1(r)&=\frac{2c^{2}r^{4}+cr^{2}-4}{r\left(1+cr^{2}\right)\left(2-cr^{2}\right)},\hspace{1cm}	g_3(r)=\frac{a\left(2-2cr^{2}-c^{2}r^{4}\right)+b\left(2cr^{2}-1\right)\left(2+cr^{2}\right)f_2(r)}{r\left(1+cr^{2}\right)\left(2-cr^{2}\right)\left(a-bf_2(r)\right)},
		\\[0.3cm]
		g_2(r)&=\frac{2a\left(2-cr^{2}\right)+b\left(3cr^{2}-4\right)f_2(r)}{r\left(2-cr^{2}\right)\left(a-bf_2(r)\right)}, \hspace{1cm} 	h_1(r)=\frac{c\left[3a\left(3+cr^{2}\right)-b\left(12+5cr^{2}\right)f_2(r)\right]}{\left(1+cr^{2}\right)^{2}\left(a-bf_2(r)\right)},\\[0.3cm]
		h_2(r)&=-\frac{6c^{2}r\left(5+cr^{2}\right)}{\left(1+cr^{2}\right)^{3}}+\frac{bc^{2}r\left(19+5cr^{2}-2c^{2}r^{4}\right)f_2(r)}{\left(1+cr^{2}\right)^{3}\left(2-cr^{2}\right)\left(a-bf_2(r)\right)}+\frac{3b^{2}cr\left(3+2cr^{2}\right)}{\left(1+cr^{2}\right)^{2}\left(a-bf_2(r)\right)^{2}},
		\\[0.3cm]
		h_3(r)&=\frac{6c^{2}\left(3c^{2}r^{4}+22cr^{2}-5\right)}{\left(1+cr^{2}\right)^{4}}+\frac{2bc^{2}\left(19-80cr^{2}+27c^{2}r^{4}+16c^{3}r^{6}-2c^{4}r^{8}\right)f_2(r)}{\left(1+cr^{2}\right)^{4}\left(2-cr^{2}\right)^{2}\left(a-bf_2(r)\right)}\\
		&+\frac{6b^{2}c\left(3-14cr^{2}-3c^{2}r^{4}+2c^{3}r^{6}\right)}{\left(1+cr^{2}\right)^{3}\left(2-cr^{2}\right)\left(a-bf_2(r)\right)^{2}}-\frac{6b^{3}c^{2}r^{2}\left(3+2cr^{2}\right)f_2(r)}{\left(1+cr^{2}\right)^{2}\left(2-cr^{2}\right)\left(a-bf_2(r)\right)^{3}},
		\\[0.3cm]
		f_1(r)&=\frac{2-cr^2}{2+2cr^2}, \hspace{1.5cm} f_2(r)=\sqrt{\frac{6-3cr^2}{c}}.
	\end{split}
\end{equation}
\par
\subsection{Hayward exterior}\label{sec:exterior}

Far from the stellar distribution, spacetime is expected to approach the Minkowski spacetime. Because of this, the interior geometry must be matched with a suitable exterior spacetime. In GR, Birkhoff's theorem settles the question: the exterior is the Schwarzschild metric \citep{h8}. This metric is perfectly regular in the region $r\ge R$ where it is actually used, and its singularity at $r=0$ is never part of a stellar spacetime. Things are different in $\mathcal{F}(\mathcal{R})$ gravity, where the additional scalar degree of freedom carried by $\mathcal{F}_{\mathcal{R}}$ obeys the trace of Eq.~(\ref{6e}),
\begin{equation}\label{eq:trace}
3\,\Box\mathcal{F}_{\mathcal{R}}+\mathcal{F}_{\mathcal{R}}\mathcal{R}-2\mathcal{F}=8\pi T
\quad\Longrightarrow\quad
\Box\mathcal{R}-\frac{\mathcal{R}}{6\alpha}=\frac{8\pi T}{6\alpha}
\qquad\bigl(\mathcal{F}=\mathcal{R}+\alpha\mathcal{R}^{2}\bigr).
\end{equation}
The Ricci scalar is thus a propagating field, with Compton wavelength $\lambda_\alpha=\sqrt{6\alpha}$ (about $0.8$~km for $\alpha=0.1$~km$^{2}$), and it need not vanish outside matter. Birkhoff's theorem is lost, and self-consistent stellar solutions of the theory do have exteriors that differ from Schwarzschild close to the surface \citep{c4,Yazadjiev2014,CasadoTurrion2023}. Which exterior to use is thus an assumption of the model, and its consequences need to be quantified.
In the present work, we consider the Hayward metric for the exterior region:
\begin{equation}\label{22e}
	ds_2^2=-C(r)\, dt^2+C^{-1}(r)\, dr^2+r^2(d\theta^2+\sin^2\theta\, d\phi^2),
\end{equation}
where $C(r)=1-2Mr^{2}/(r^{3}+2\ell^{2}M)$ and $\ell$ is the Hayward length.  The Hayward metric can be characterized by the following metric function

\begin{equation}\label{23e}
	C(r)=1-\frac{2M}{r}\left(1+\frac{2M\ell^2}{r^3}\right)^{-1}
	=1-\frac{2M}{r}+\frac{4M^2\ell^2}{r^4}+O\left(\frac{1}{r^7}\right).
\end{equation}
here, the term $1-\frac{2M}{r}$ corresponds to the classical Schwarzschild part and fixes the ADM mass $M$. The term $+4M^{2}\ell^{2}/r^{4}$ is the leading Hayward correction. It is positive and therefore repulsive. We relate $\ell$ to the mass through the constant ratio $\beta=\ell/M>0$. At $r=R$ this correction divided by $2M/R$ equals $2\beta^{2}(M/R)^{3}$ [cf.\ Eq.~(\ref{eq:betashift})], which is only about $3\times10^{-4}$ for J0740+6620 with $\beta=0.1$, and since it decays as $r^{-4}$ the metric remains asymptotically flat. For $\ell=0$ we get back the Schwarzschild metric.\\
The Hayward metric \citep{h16} is used here for three reasons. First, Eq.~(\ref{23e}) shows that it is a minimal one-parameter deformation of Schwarzschild which keeps the ADM mass and asymptotic flatness. Second, it has a known field-theoretic source, GR coupled to nonlinear electrodynamics with a magnetic charge \citep{FanWang2016}, which has the effective energy density $\rho_H=3M^{2}\ell^{2}/[2\pi(r^{3}+2M\ell^{2})^{2}]>0$ and the radial stress $p_r=-\rho_H$. Third, and this is what matters most here, its effect on surface quantities can be found in closed form. Expanding to first nonvanishing order in $\beta$, with $u=M/R$, we find that the constant $c$ [Eq.~(\ref{eq:ac})] and the surface redshift [Eq.~(\ref{eq:zs})] shift by
\begin{equation}\label{eq:betashift}
\frac{\delta c}{c}=-\frac{2\beta^{2}u^{3}}{1-\tfrac{4}{3}u},\qquad
\delta z_s=-\frac{2\beta^{2}u^{4}}{(1-2u)^{3/2}},
\end{equation}
which tell us exactly when the exterior matters. The regular de~Sitter core of the Hayward geometry is located at $r\lesssim(2\beta^{2})^{1/3}M$, deep inside the star, and plays no role in our model, since we use the Hayward metric as a controlled deformation of the exterior and not for removing a singularity. We can also estimate $\ell$ assuming that it really comes from nonlinear electrodynamics. At the surface of PSR~J0740+6620, the required field energy density would then equal that of a magnetic field $B_{\rm eq}\simeq6\times10^{16}\,(\beta/0.1)$~G. A field of magnetar strength, $10^{15}$~G, would give no more than $\beta\simeq2\times10^{-3}$. Any physically sourced deformation is therefore tiny, and larger values of $\beta$ should be understood as phenomenological.

\subsection{Junction conditions in quadratic $\mathcal{F}(\mathcal{R})$ gravity}\label{sec:junction}
Let $\Sigma$ denote the surface $r=R$, with outward unit normal $n=B^{-1}\partial_r$, and write $[X]=X^{+}-X^{-}$ for the jump of a quantity across it ($+$ outside, $-$ inside). When $\mathcal{F}''\neq0$, any matching has to satisfy \citep{Deruelle2008,Senovilla2013}
\begin{equation}\label{eq:jc-always}
[h_{ab}]=0,\qquad[K^{\mu}{}_{\mu}]=0,
\end{equation}
where $h_{ab}$ and $K_{ab}$ are the first and second fundamental forms of $\Sigma$. If no thin shell is to appear, one needs in addition $[K_{ab}]=0$, $[\mathcal{R}]=0$ and $[n^{\mu}\nabla_{\mu}\mathcal{R}]=0$. For the metrics (\ref{13e}) and (\ref{22e}) we have $K^{\theta}{}_{\theta}=1/(RB)$ and $K^{t}{}_{t}=A'/(AB)$ on the inside, and $K^{\theta}{}_{\theta}=\sqrt{C}/R$ and $K^{t}{}_{t}=C'/(2\sqrt{C})$ on the outside. Continuity of $h_{ab}$ gives $A^{2}(R)=C(R)$ and $B^{2}(R)=C^{-1}(R)$, and therefore
\begin{equation}\label{eq:ac}
c=\frac{4M}{3R^{3}-4MR^{2}+6\beta^{2}M^{3}},\qquad a=\sqrt{C(R)}+b\,f_2(R).
\end{equation}
The component $K^{\theta}{}_{\theta}$ is then automatically continuous, so $[K^{\mu}{}_{\mu}]=0$ reduces to $[A'/A]=0$. Since $A'(r)=3br/f_2(r)$, this fixes $b$ in closed form,
\begin{equation}\label{eq:b}
b=\frac{f_2(R)\,C'(R)}{6R\sqrt{C(R)}}
 =\frac{f_2(R)\,M\,(R^{3}-4\ell^{2}M)}{3\,(R^{3}+2\ell^{2}M)^{2}\sqrt{C(R)}} .
\end{equation}
Note that none of the constants $a$, $b$ and $c$ depends on $\alpha$. The conditions that fix them involve only the metric, and $\alpha$ enters only through the matter variables of Eq.~(\ref{15e}). In GR with $\ell=0$, Eq.~(\ref{eq:b}) is equivalent to $p_r(R)=0$. In every other case it is not, and imposing $p_r(R)=0$, as earlier work did, violates the mandatory condition (\ref{eq:jc-always}). With $b$ taken from Eq.~(\ref{eq:b}), the radial pressure at $\Sigma$ follows in closed form from the $rr$ component of Eq.~(\ref{15e}),
\begin{equation}\label{eq:prR}
8\pi p_r(R)=-8\pi\rho_H(R)\bigl[1+2\alpha\mathcal{R}^{-}\bigr]+2\alpha C\,(\mathcal{R}^{-})'\Bigl(\frac{C'}{2C}+\frac{2}{R}\Bigr)+\frac{\alpha}{2}\bigl(\mathcal{R}^{-}\bigr)^{2},
\end{equation}
where $\mathcal{R}^{-}$ and $(\mathcal{R}^{-})'$ are the interior Ricci scalar and its derivative at $R$, $C$ and $C'$ are evaluated at $R$, and $\rho_H$ is the energy density of the Hayward source (Sec.~\ref{sec:exterior}). In GR $p_r(R)=-\rho_H(R)$, which is simply the radial tension of the exterior source. For every configuration we consider, the $\alpha$-terms push $p_r(R)$ to even more negative values. For PSR~J0740+6620 with $\alpha=0.1$~km$^{2}$ and $\beta=0.1$, for example, $p_r(R)=-1.3\times10^{-7}$~km$^{-2}$, or $0.15\%$ of the central pressure. For all pulsars of Table~\ref{tab:obs}, $p_r$ becomes zero at a radius $r_0$ which is 17--67~m below $\Sigma$, and $|p_r(R)|$ is between 0.15\% and 0.85\% of $p_c$. We require $|p_r(R)|\le0.01\,p_c$, and this condition gives bounds on $\alpha$ and $\ell$ (Sec.~\ref{sec:obs-bounds}). The remaining conditions, for $\mathcal{R}$ and its normal derivative, cannot be fulfilled. Inside the star, $\mathcal{R}(R)$ is fixed by the metric and is positive, of order $8\pi\rho_s$, whereas outside $\mathcal{R}_{H}(r)=24M^{2}\ell^{2}(4M\ell^{2}-r^{3})/(r^{3}+2M\ell^{2})^{3}$, which is negative for $r^{3}>4M\ell^{2}$. Quadratic $\mathcal{F}(\mathcal{R})$ gravity tolerates a discontinuous $\mathcal{R}$, but the jump acts as a source concentrated on $\Sigma$ \citep{Senovilla2013,Reina2016}. This source consists of a shell with
\begin{equation}\label{eq:shell}
8\pi\tau_{ab}=-\bigl[1+\alpha(\mathcal{R}^{+}+\mathcal{R}^{-})\bigr][K_{ab}]
+\alpha\Bigl\{2\,n^{\mu}[\nabla_{\mu}\mathcal{R}]\,h_{ab}-[\mathcal{R}]\bigl(K^{+}_{ab}+K^{-}_{ab}\bigr)\Bigr\},
\end{equation}
an external tension $8\pi\tau=2\alpha[\mathcal{R}]K^{\mu}{}_{\mu}$, and a gravitational double layer whose strength is set by $2\alpha[\mathcal{R}]h_{ab}$. Because $[K_{ab}]=0$ by Eq.~(\ref{eq:b}), the shell is a pure $\alpha$ effect. For PSR~J0740+6620 with $\alpha=0.1$~km$^{2}$ and $\beta=0.1$ we find $[\mathcal{R}]=-6.4\times10^{-3}$~km$^{-2}$, a surface energy density $\sigma=-\tau^{t}{}_{t}=-2.1\times10^{-6}$~km$^{-1}$ and a surface pressure $\mathcal{P}=\tau^{\theta}{}_{\theta}=3.6\times10^{-6}$~km$^{-1}$. Since $\sigma<0$ for all $\alpha>0$, the weak energy condition is violated on the shell. Its mass, $4\pi R^{2}|\sigma|\simeq0.14\%$ of $M$, is of the same order as the changes produced by $\alpha$ in the interior. To remove the shell completely one would need an interior ansatz with more free parameters or an exterior that solves the $\mathcal{R}+\alpha\mathcal{R}^{2}$ field equations, and we leave this for future work.

 \section{\texorpdfstring{Physical Admissibility of the Interior}{Physical Admissibility of the Interior}}\label{4s}
\begin{figure}[hbt!]
	\begin{minipage}{0.48\textwidth}
		\includegraphics[width=\linewidth]{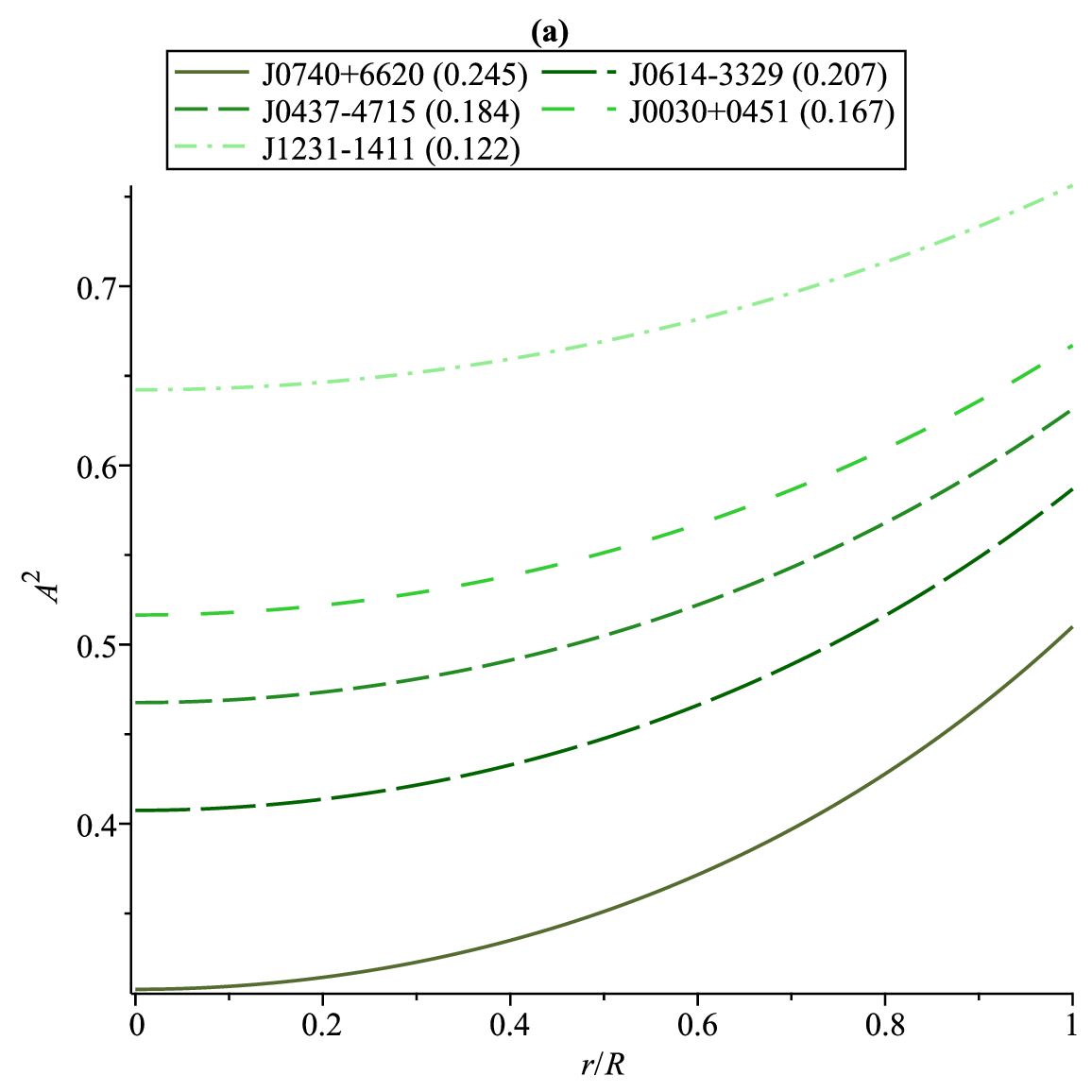}
	\end{minipage}
	\hfill
	\begin{minipage}{0.48\textwidth}
		\includegraphics[width=\linewidth]{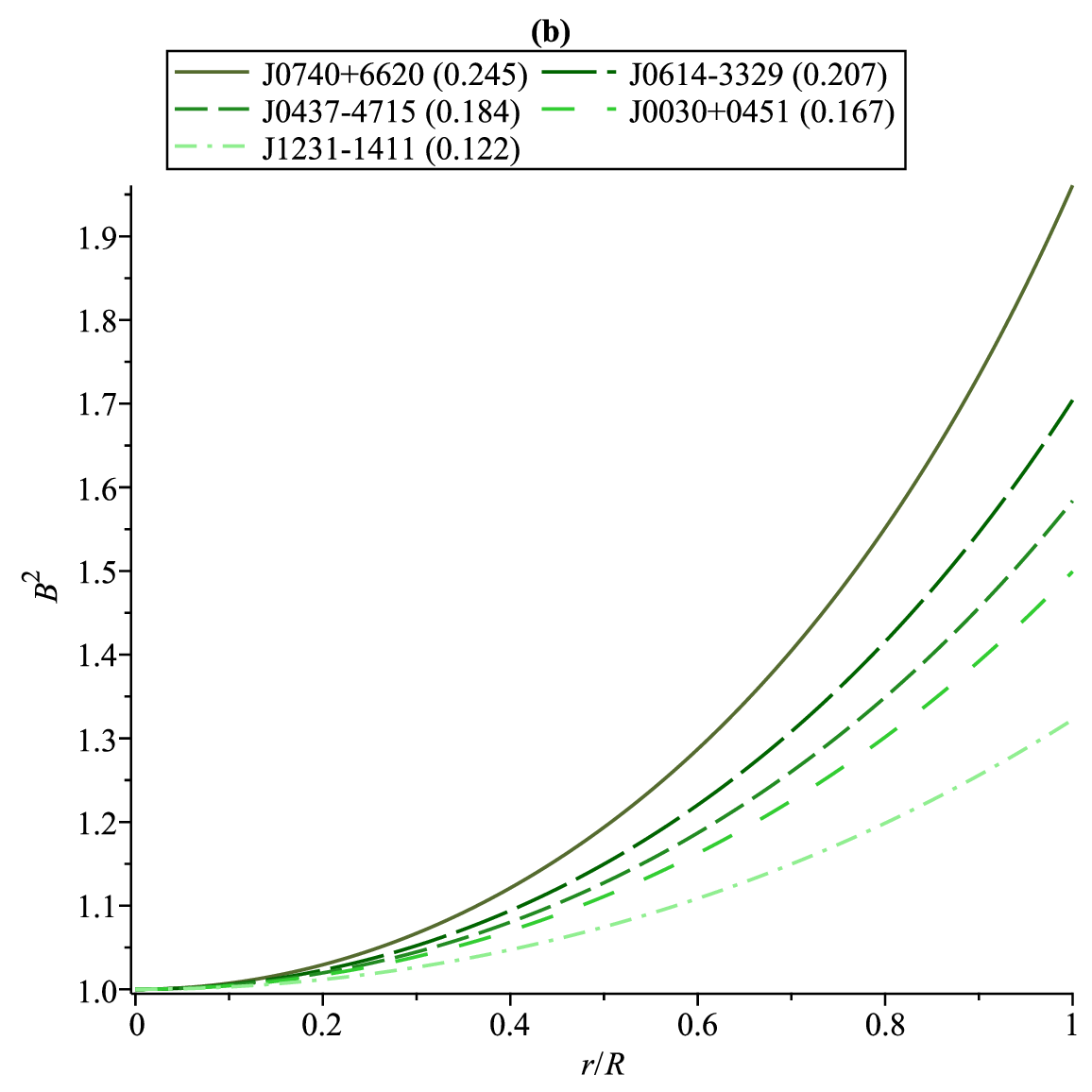}
	\end{minipage}
	\caption{Metric potentials (a) $A^{2}(r)$ and (b) $B^{2}(r)$ as functions of $r/R$ for the pulsars of Table~\ref{tab:obs} ($\alpha=0.1$~km$^{2}$, $\beta=0.1$). Darker lines correspond to more compact stars ($M/R$ is given in the legend); Figs.~\ref{fig:matter}--\ref{fig:gamma} follow the same convention.}
	\label{fig:metric}
\end{figure}

\begin{figure}[hbt!]
	\begin{minipage}{0.32\textwidth}
		\includegraphics[width=\linewidth]{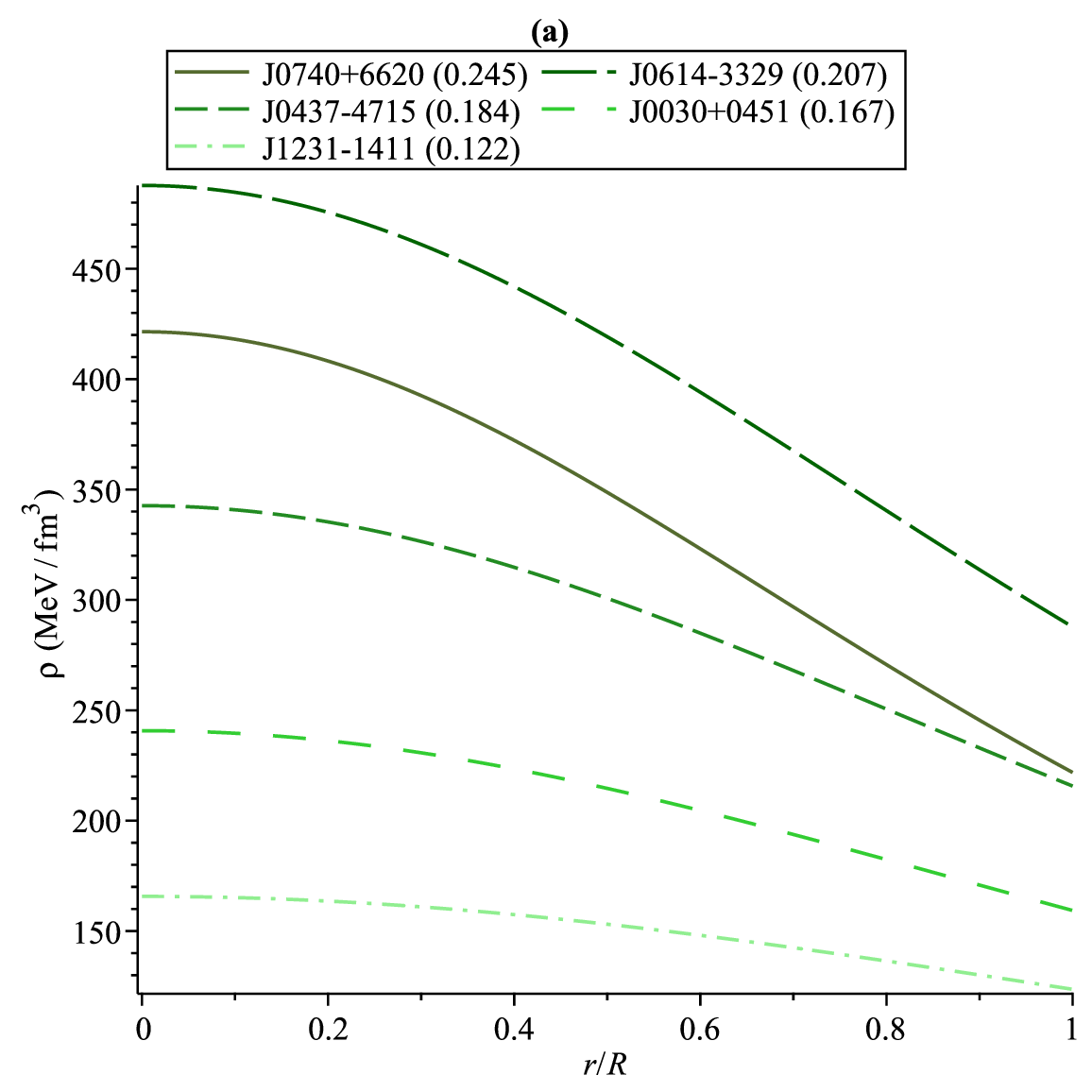}
	\end{minipage}
	\hfill
	\begin{minipage}{0.32\textwidth}
		\includegraphics[width=\linewidth]{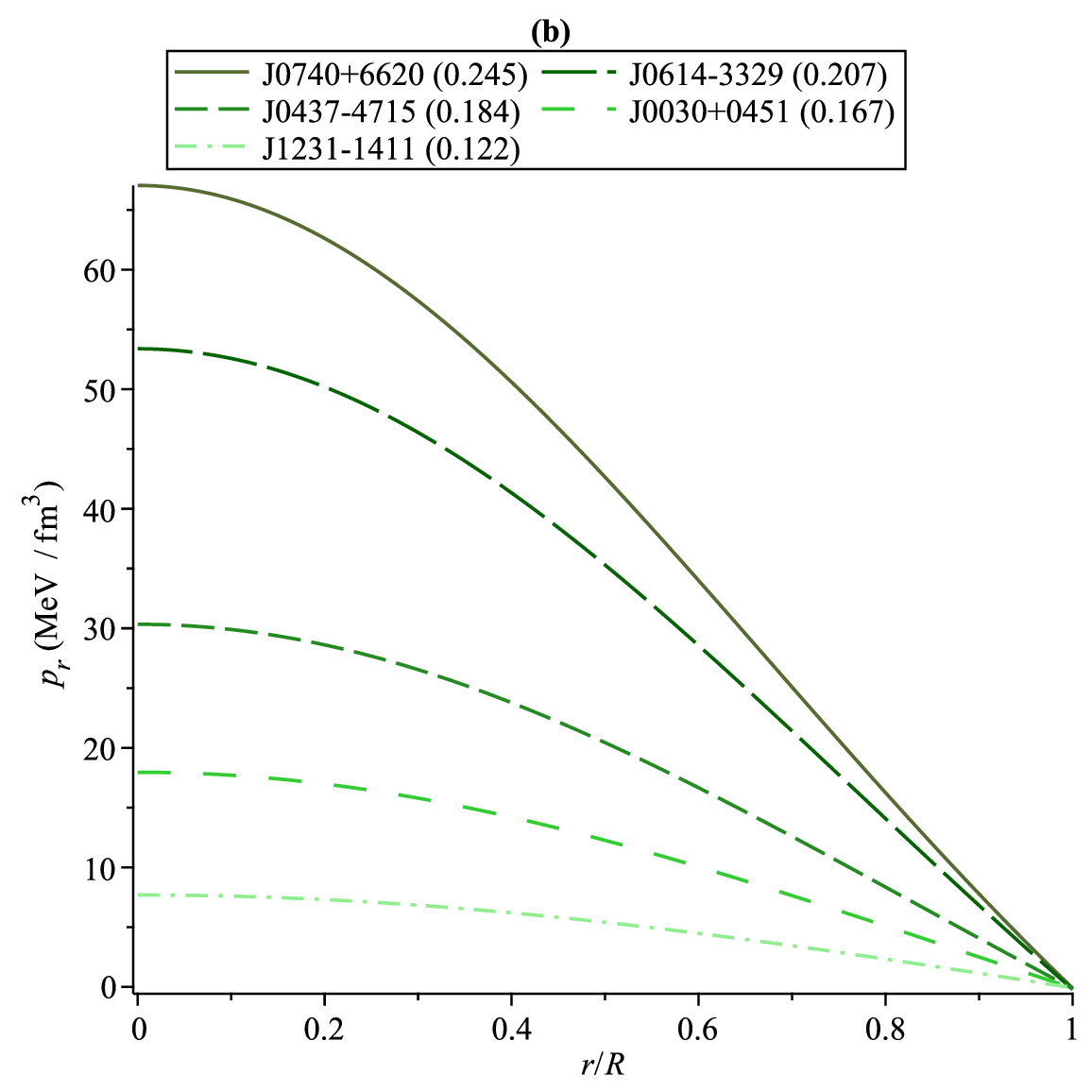}
	\end{minipage}
	\hfill
	\begin{minipage}{0.32\textwidth}
		\includegraphics[width=\linewidth]{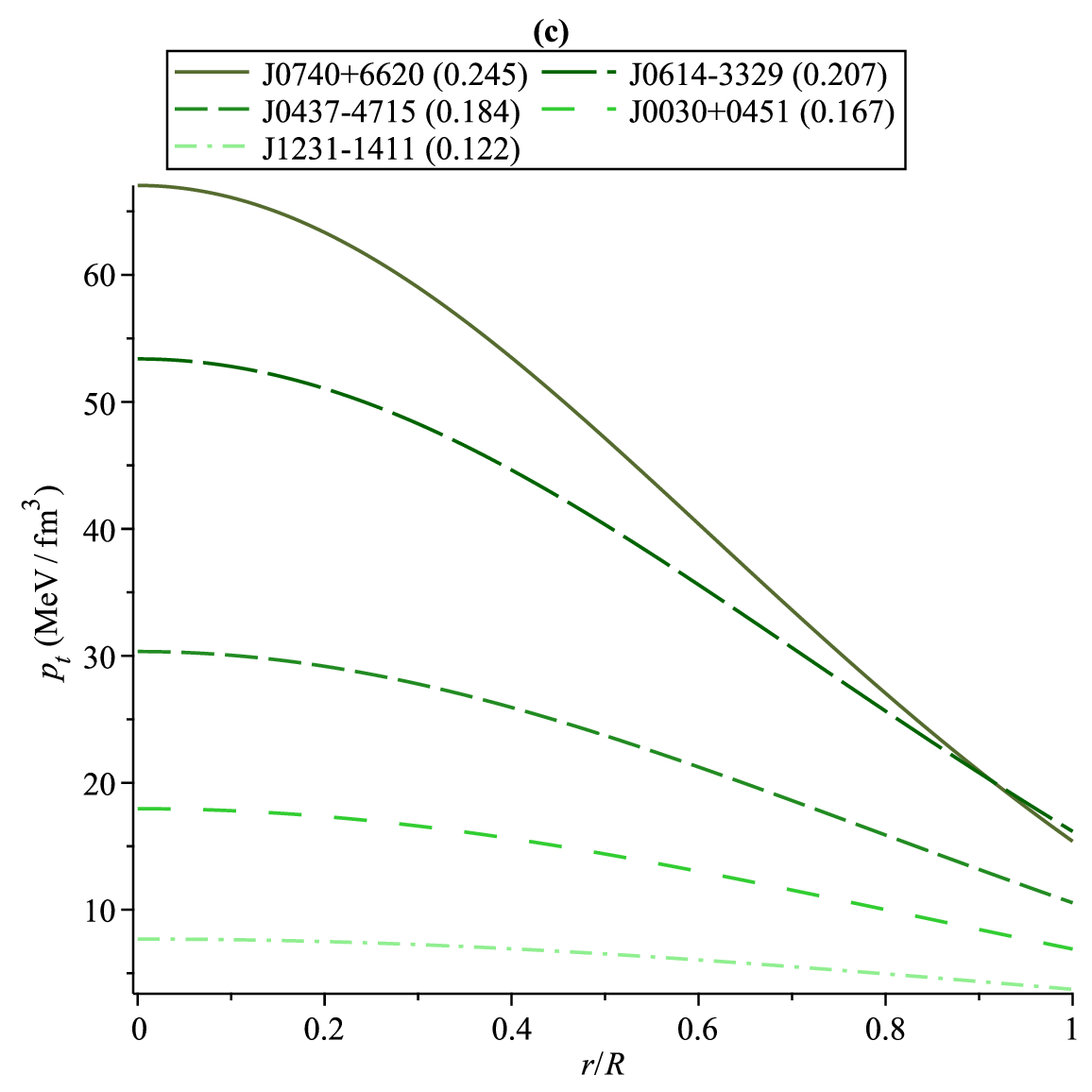}
	\end{minipage}
	\caption{(a) Energy density, (b) radial pressure and (c) tangential pressure as functions of $r/R$. The radial pressure vanishes at $r_0$, 17--67~m below the surface (Table~\ref{tab:props}).}
	\label{fig:matter}
\end{figure}

\begin{figure}[hbt!]
	\begin{minipage}{0.48\textwidth}
		\includegraphics[width=\linewidth]{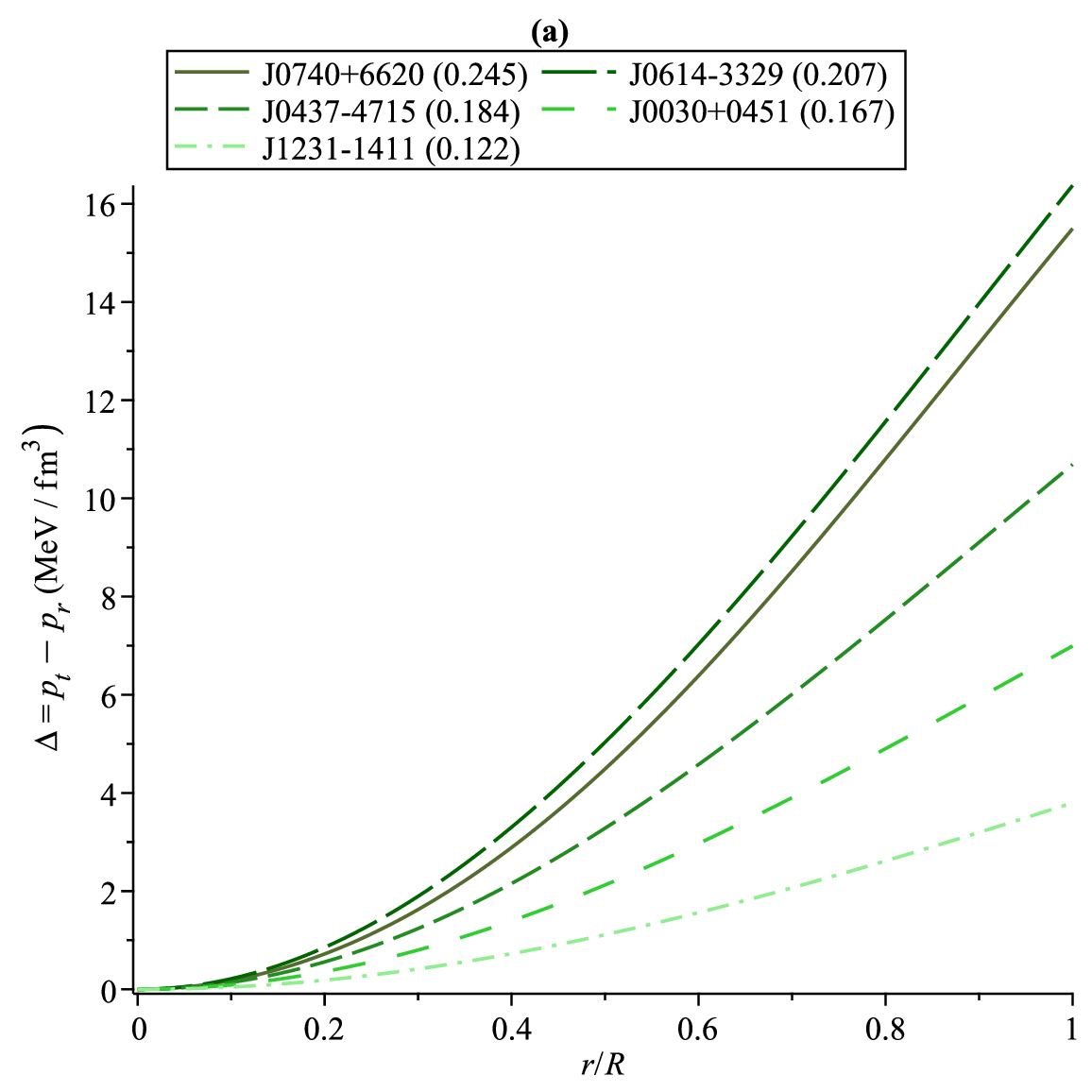}
	\end{minipage}
	\hfill
	\begin{minipage}{0.48\textwidth}
		\includegraphics[width=\linewidth]{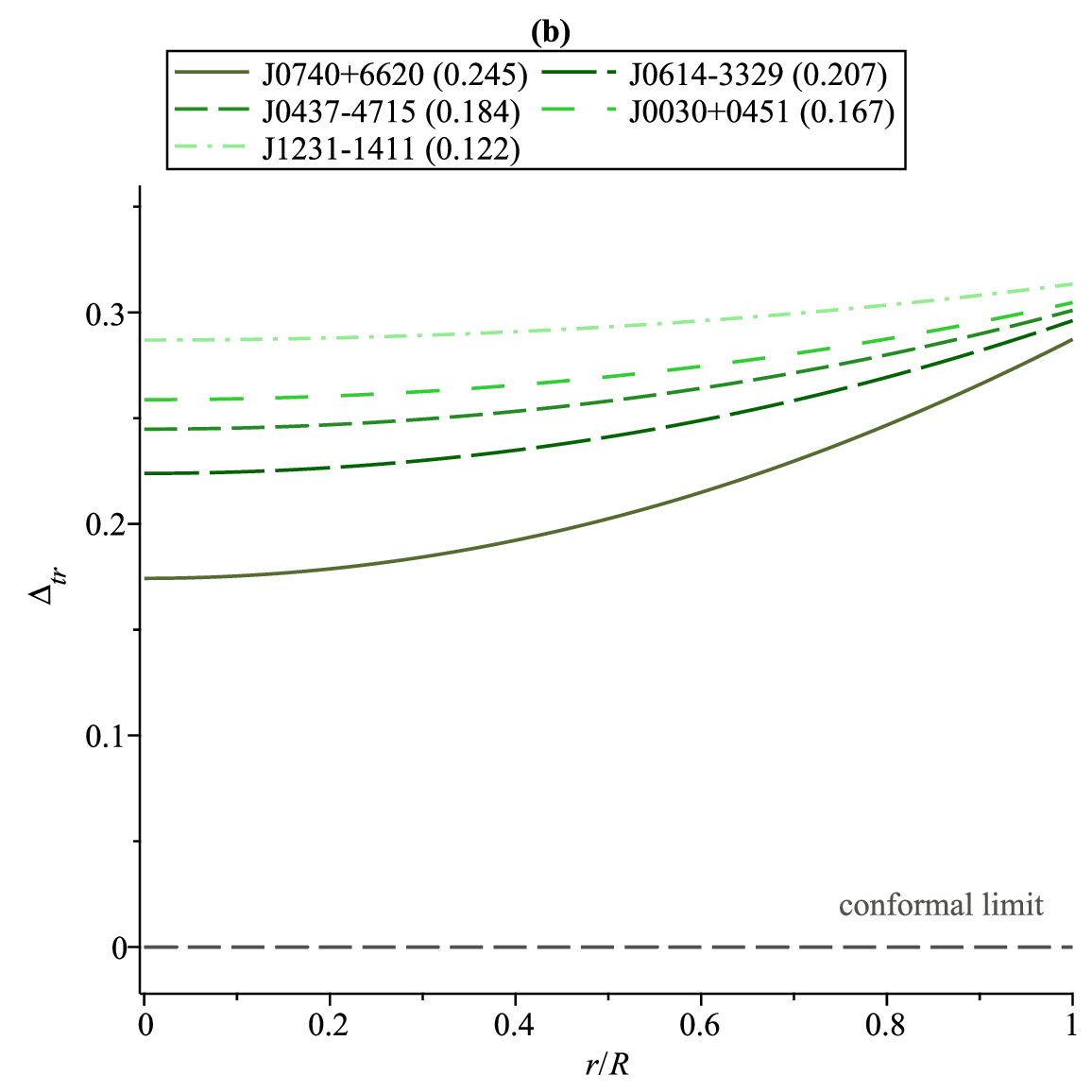}
	\end{minipage}
	\caption{(a) Anisotropy $\Delta=p_t-p_r$ and (b) normalized trace anomaly $\Delta_{\rm tr}=1/3-(p_r+2p_t)/(3\rho)$ as functions of $r/R$; the dashed line marks the conformal limit $\Delta_{\rm tr}=0$.}
	\label{fig:aniso}
	\end{figure}

\begin{figure}[hbt!]
		\begin{minipage}{0.48\textwidth}
			\includegraphics[width=\linewidth]{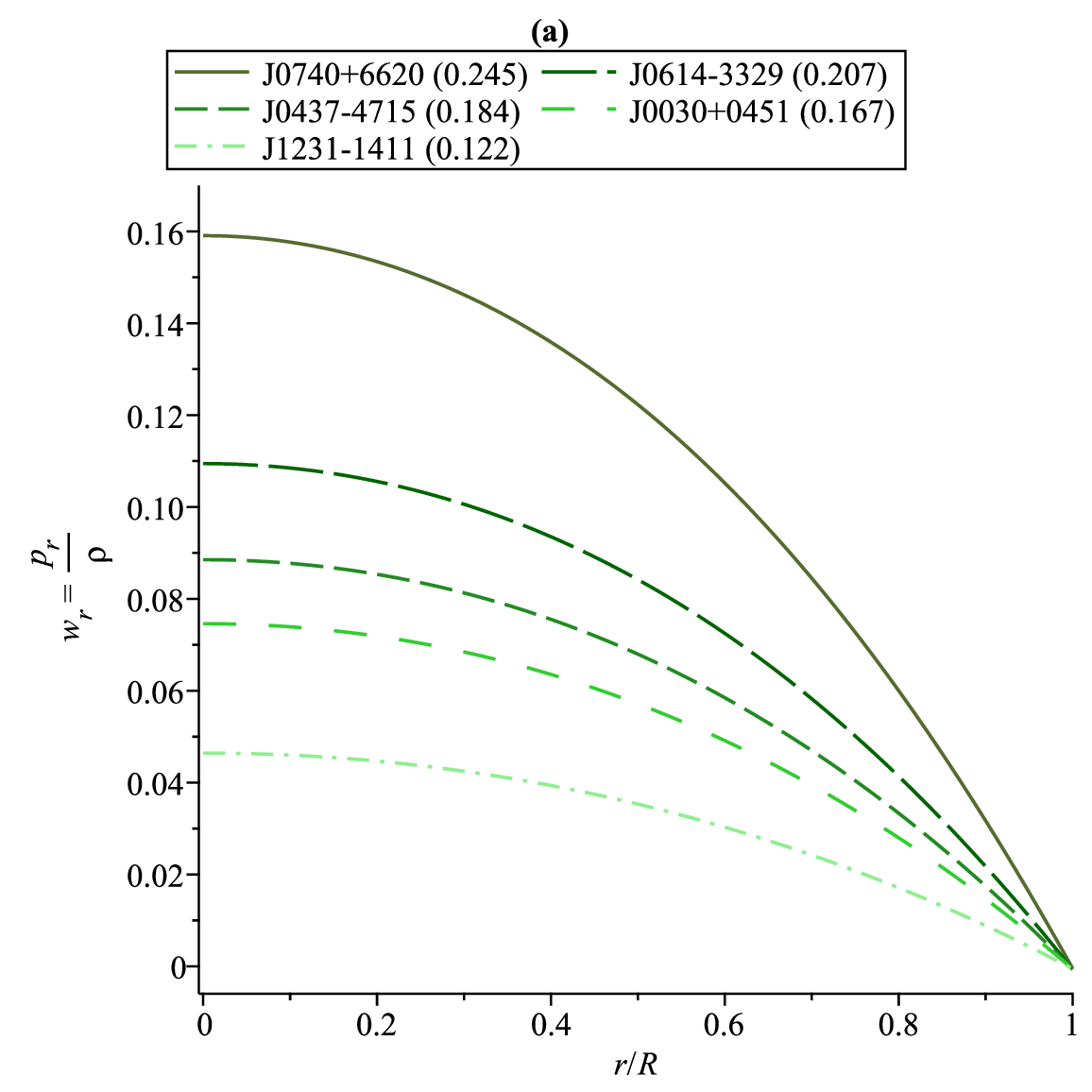}
		\end{minipage}
		\hfill
		\begin{minipage}{0.48\textwidth}
			\includegraphics[width=\linewidth]{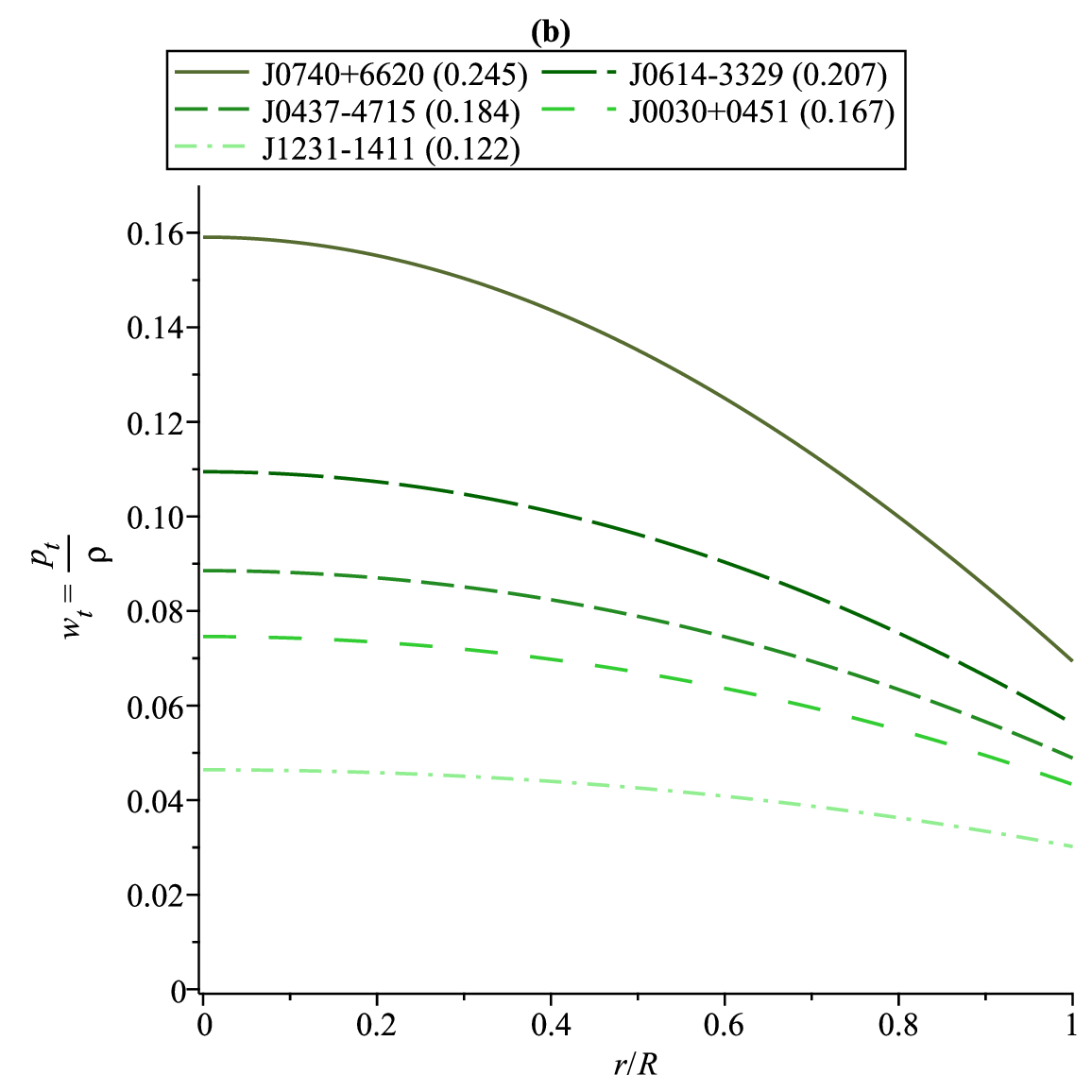}
		\end{minipage}
	     \caption{Equation-of-state ratios (a) $w_r=p_r/\rho$ and (b) $w_t=p_t/\rho$ as functions of $r/R$. Both stay below $0.16$, so the dominant energy condition is satisfied with a wide margin.}
	     \label{fig:eos}
	\end{figure}

\begin{figure}[hbt!]
	\begin{minipage}{0.32\textwidth}
		\includegraphics[width=\linewidth]{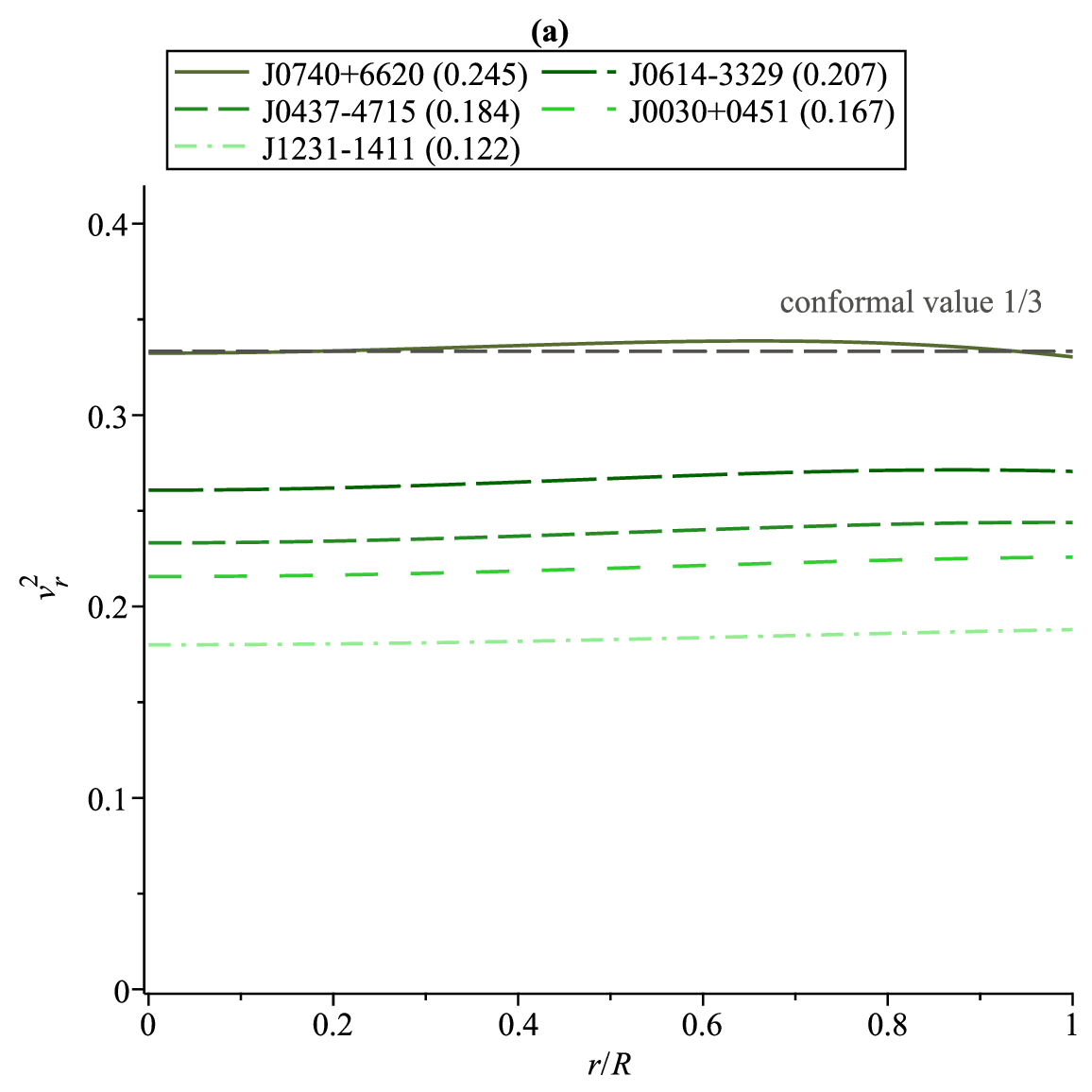}
	\end{minipage}
	\hfill
	\begin{minipage}{0.32\textwidth}
		\includegraphics[width=\linewidth]{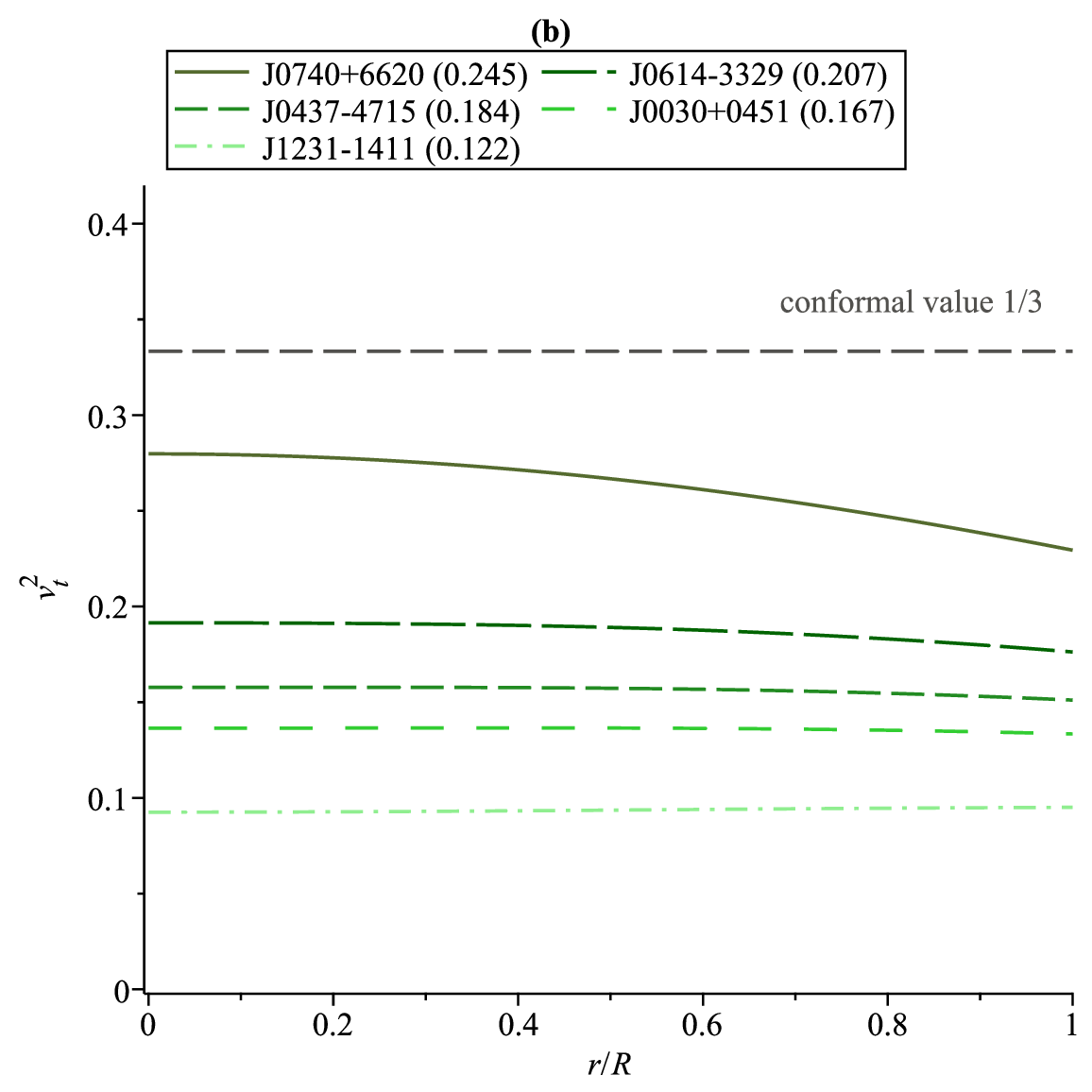}
	\end{minipage}
	\hfill
	\begin{minipage}{0.32\textwidth}
		\includegraphics[width=\linewidth]{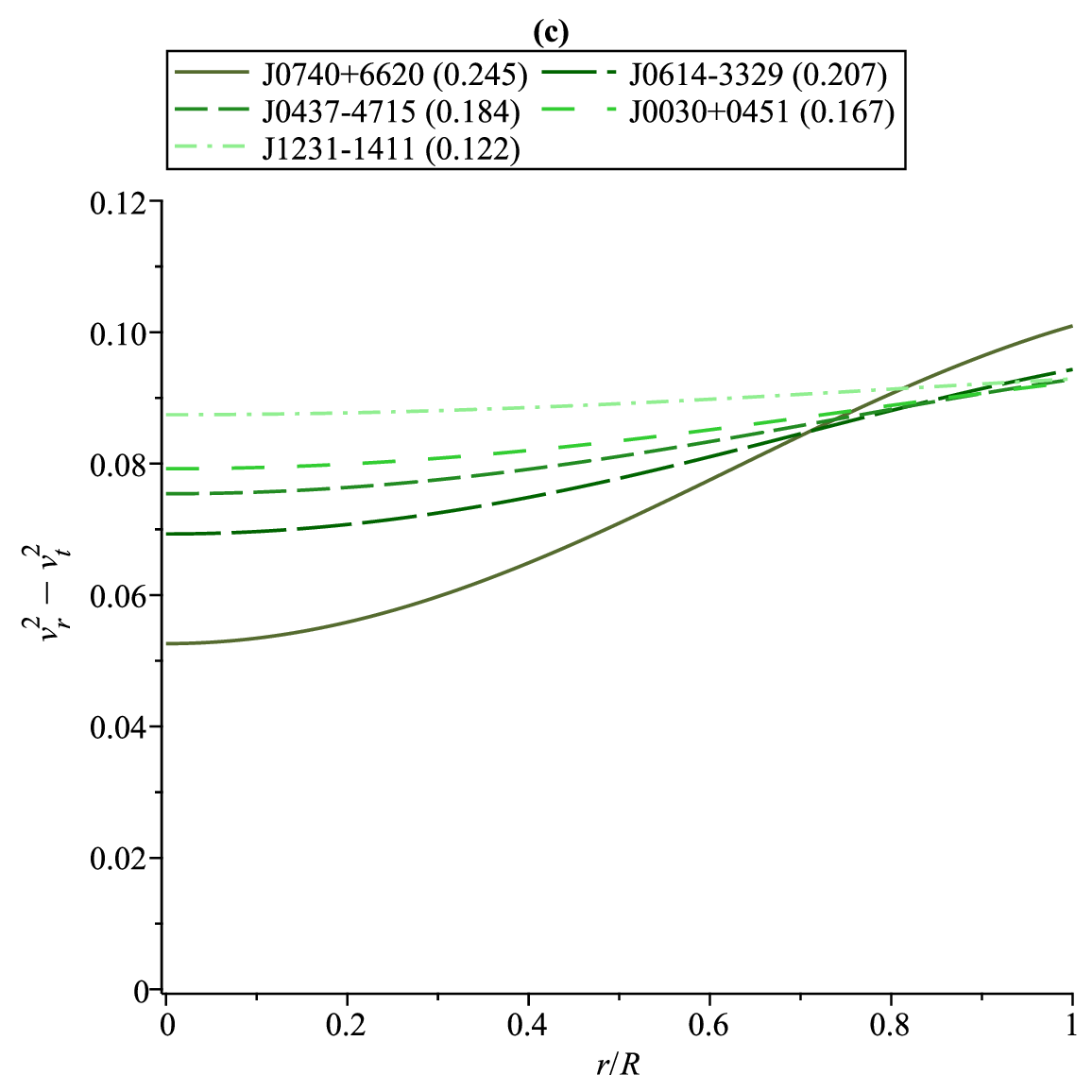}
	\end{minipage}
     \caption{(a) Radial and (b) tangential sound speeds and (c) the cracking indicator $v_r^{2}-v_t^{2}$ as functions of $r/R$. Dashed lines mark the conformal value $1/3$ in (a) and (b); in (c) the indicator never leaves the interval $[0,1]$.}
\label{fig:sound}
\end{figure}

\begin{figure}[hbt!]
		\begin{minipage}{0.48\textwidth}
			\includegraphics[width=\linewidth]{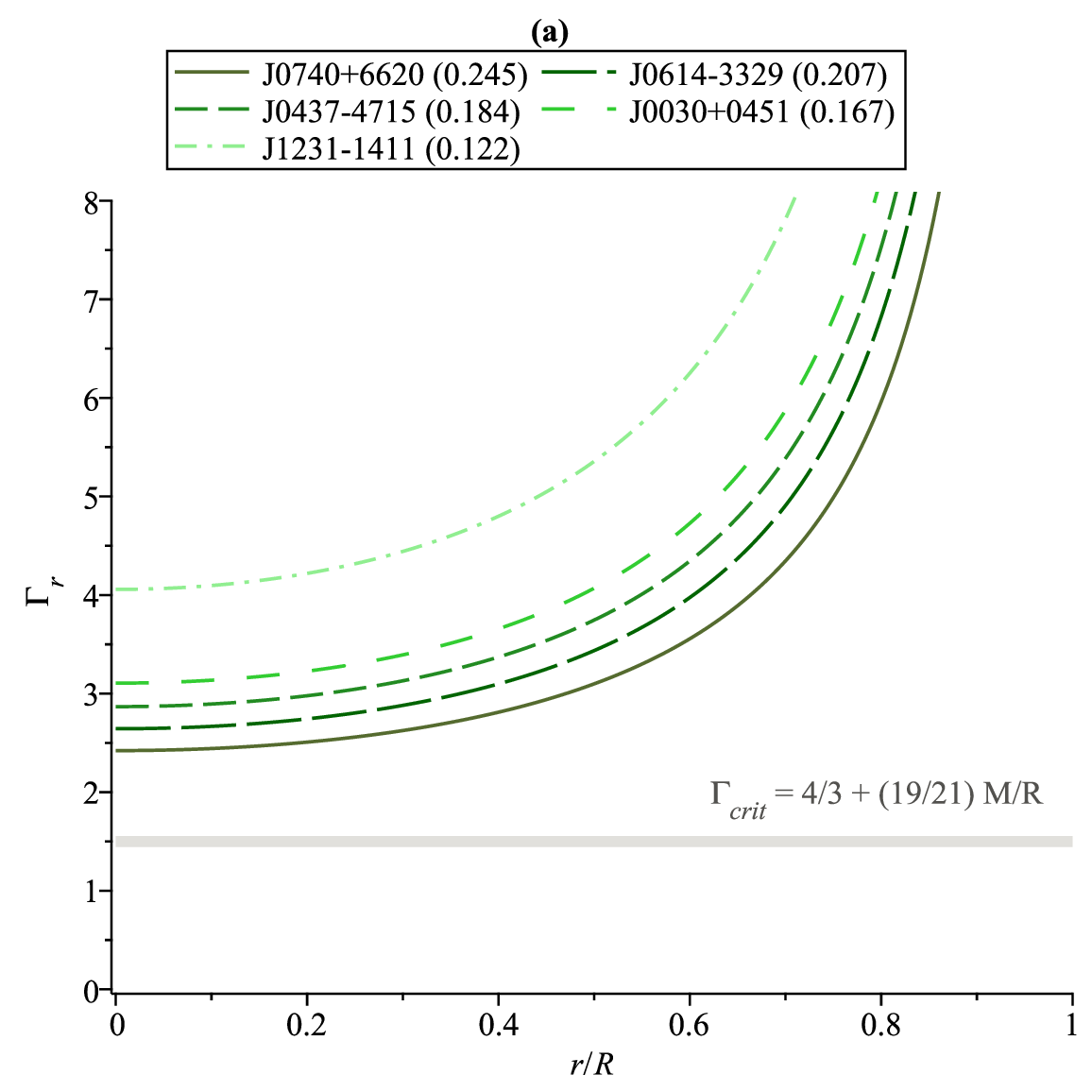}
		\end{minipage}
		\hfill
		\begin{minipage}{0.48\textwidth}
			\includegraphics[width=\linewidth]{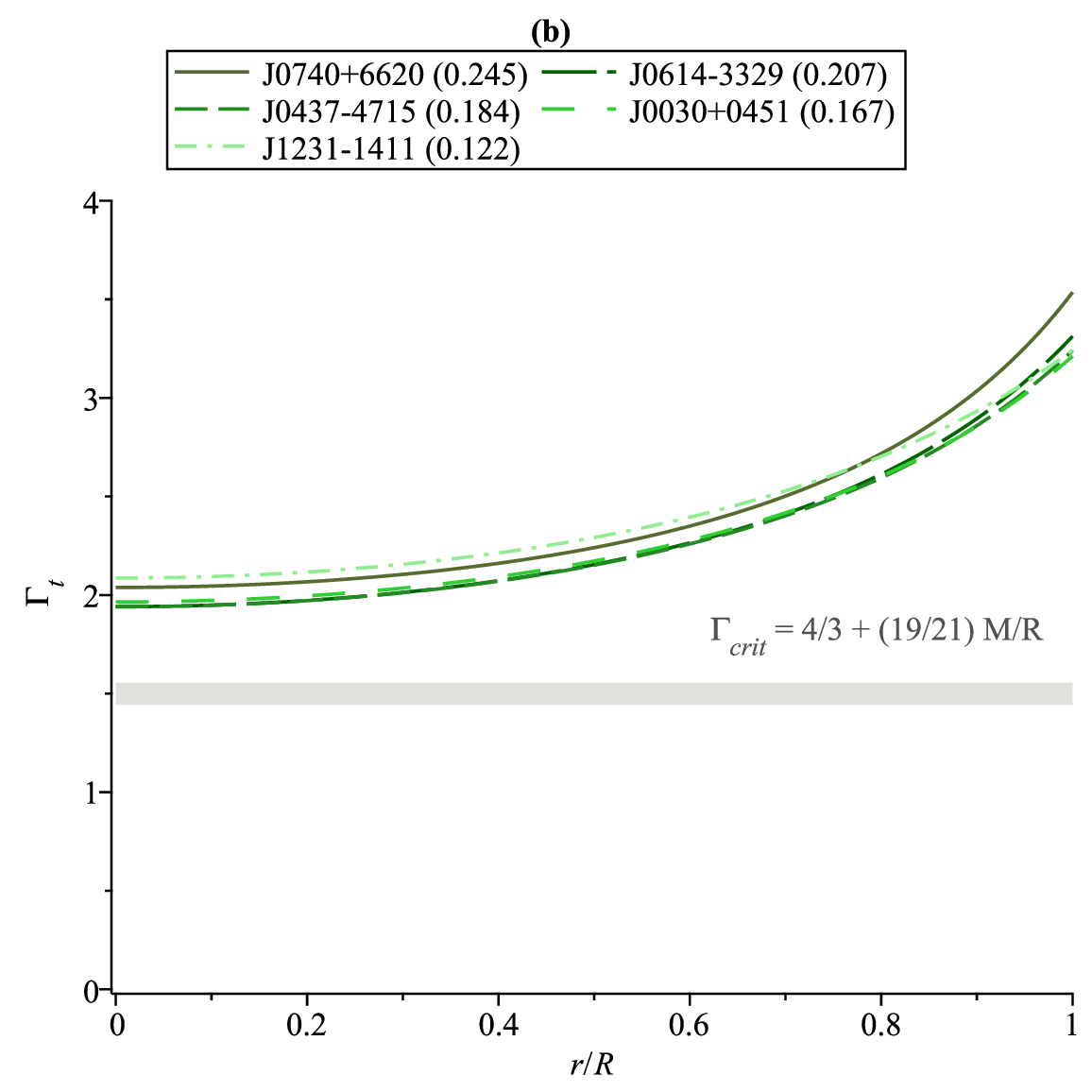}
		\end{minipage}
		\caption{Adiabatic indices (a) $\Gamma_r$ and (b) $\Gamma_t$ as functions of $r/R$. The gray band covers the values of $\Gamma_{\rm crit}=4/3+(19/21)M/R$ for the five pulsars.}
		\label{fig:gamma}
	\end{figure}

The tests can be grouped into three levels. The first level is consistency. Here the field equations (\ref{15e}) are satisfied by construction, which makes the conservation law (\ref{30e}) an identity \citep{ko}, and the junction conditions of Sec.~\ref{sec:junction} are imposed explicitly. At the second level we test whether the matter is locally admissible, i.e.\ whether $\rho$, $p_r$ and $p_t$ are positive and decrease outward (for $p_r$ below the radius $r_0$ where it vanishes), whether all four energy conditions (null, weak, strong and dominant) are satisfied, and whether the radial and tangential sound speeds obey causality, $0<v_r^{2},v_t^{2}\le1$. The third level consists of two stability indicators, the no-cracking condition $0\le v_r^{2}-v_t^{2}\le1$ \citep{P5} and the relativistic adiabatic-index criterion
\begin{equation}\label{eq:gammacrit}
\Gamma_r>\Gamma_{\rm crit}=\frac{4}{3}+\frac{19}{21}\frac{M}{R}
\end{equation}
\citep{Chandrasekhar1964,Moustakidis2017}. We carry out these tests for the five pulsars listed in Table~\ref{tab:obs} (the data are discussed in Sec.~\ref{sec:obs-data}), using $\alpha=0.1$~km$^{2}$ and $\beta=0.1$. Table~\ref{tab:consts} lists $a$, $b$ and $c$ for each pulsar and Table~\ref{tab:props} the corresponding stellar properties. In Figs.~\ref{fig:metric}--\ref{fig:gamma} the profiles are shown as functions of $r/R$ for all five stars. The same tests are applied again in Sec.~\ref{sec:obs}, where they determine the maximum mass and the bounds on $\alpha$ and $\ell$.
\begin{table}[hbt!]
\caption{Masses and radii of the five pulsars (median values with 68\% credible intervals).}
\label{tab:obs}
\centering
\begin{tabular}{p{2cm}p{2cm}p{2cm}p{2cm}c}
\toprule
Pulsar & $M/M_\odot$ & $R$ (km) & $\rho_s/\rho_0$ & Reference \\
\midrule
J0740+6620   & $2.073\pm0.069$        & $12.49^{+1.28}_{-0.88}$ & $1.46^{+0.32}_{-0.35}$ & \citet{Salmi2024J0740}\\
J0437$-$4715 & $1.418\pm0.037$        & $11.36^{+0.95}_{-0.63}$ & $1.42^{+0.25}_{-0.29}$ & \citet{Choudhury2024}\\
J0030+0451   & $1.43^{+0.20}_{-0.17}$ & $12.68^{+1.31}_{-1.04}$ & $1.05^{+0.31}_{-0.28}$ & \citet{Kini2026}\\
J0614$-$3329 & $1.44^{+0.06}_{-0.07}$ & $10.29^{+1.01}_{-0.86}$ & $1.90^{+0.52}_{-0.44}$ & \citet{Mauviard2025}\\
J1231$-$1411 & $1.04^{+0.05}_{-0.03}$ & $12.6\pm0.3$            & $0.82^{+0.07}_{-0.06}$          & \citet{Salmi2024J1231}\\
\bottomrule
\end{tabular}
\footnotetext{The radius of J1231$-$1411 depends on the assumed radius prior, so this source serves only as a cross-check. The surface energy density obtained from each pair in the model, for $\alpha=0.1$ and $\beta=0.1$ ($\rho_0=2.7\times10^{14}$~g~cm$^{-3}$).}
\end{table}

\begin{table}[hbt!]
\caption{Model constants at the central $(M,R)$ of each pulsar in Table~\ref{tab:obs}, $M_\odot=1.4766$ km.}
\label{tab:consts}
\centering
\begin{tabular}{p{3cm}p{1cm}p{1.5cm}p{1.5cm}cc}
\toprule
Pulsar & $M/M_\odot$ & $R$ (km) & $a$ & $10^{2}\,b$ (km$^{-1}$) & $10^{3}\,c$ (km$^{-2}$)\\
\midrule
J0740+6620   & 2.073 & 12.49 & 1.784399843 & 2.799786354 & 3.109986158\\
J0437$-$4715 & 1.418 & 11.36 & 1.986094618 & 2.671140234 & 2.524395300\\
J0030+0451   & 1.43  & 12.68 & 2.041382615 & 2.274915045 & 1.774873448\\
J0614$-$3329 & 1.44  & 10.29 & 1.914458901 & 3.121809286 & 3.590741323\\
J1231$-$1411 & 1.04  & 12.6  & 2.173927990 & 1.958945023 & 1.222142518\\
\bottomrule
\end{tabular}
\footnotetext{ We give ten significant figures because the small radial stress at the surface, $p_r(R)$, is sensitive to rounding.}
\end{table}

\begin{table}[hbt!]
\caption{Properties of the configurations at the central $(M,R)$ of each pulsar in Table~\ref{tab:obs}.}
\label{tab:props}
\centering
\footnotesize
\setlength{\tabcolsep}{4.5pt}
\begin{tabular}{@{}lccccccc@{}}
\toprule
Pulsar & $u$ & $z_s$ & $\rho_c$ (MeV\,fm$^{-3}$) & $p_c$ (MeV\,fm$^{-3}$) & $\max v_r^{2}$ & $\Gamma_{r,\min}$ ($\Gamma_{\rm crit}$) & $R-r_0$ (m)\\
\midrule
J0740+6620   & 0.245 & 0.400 & 421 & 67.0 & 0.339 & 2.42 (1.56) & 17\\
J0437$-$4715 & 0.184 & 0.258 & 343 & 30.3 & 0.244 & 2.87 (1.50) & 38\\
J0030+0451   & 0.167 & 0.224 & 241 & 17.9 & 0.226 & 3.11 (1.48) & 44\\
J0614$-$3329 & 0.207 & 0.305 & 488 & 53.4 & 0.271 & 2.64 (1.52) & 30\\
J1231$-$1411 & 0.122 & 0.150 & 166 & 7.7  & 0.188 & 4.06 (1.44) & 67\\
\bottomrule
\end{tabular}
\footnotetext{ Here $u=M/R$ is the compactness, $z_s$ the surface redshift, $\rho_c$ and $p_c$ the central energy density and pressure, and $\Gamma_{\rm crit}$ the critical index of adiabatic-index, $r_0$ is the radius at which $p_r$ vanishes.}
\end{table}
\par
\noindent The radial behavior of the metric potentials of the considered stellar models is illustrated in Fig.~\ref{fig:metric}. Both metric functions remain finite throughout the interior region, indicating the absence of singularities. Also, the functions $A^2(r)$ and $B^2(r)$ increase monotonically, while the condition $B^2(r)=1$ at the center is satisfied. Both potentials therefore behave as required.
	\par \noindent The distributions of the energy density $\rho$ and the radial and tangential pressures $(p_r,p_t)$ are illustrated in Fig.~\ref{fig:matter}. All three are regular and positive; the only exception is $p_r$, which crosses zero at $r_0$, 17--67~m below the matching surface, as the junction conditions of Sec.~\ref{sec:junction} require. Their maximum values occur at the center and decrease monotonically toward the boundary, reflecting a physically realistic matter distribution. The gradients $d\rho/dr$, $dp_r/dr$ and $dp_t/dr$ are negative everywhere. The central energy densities lie between 166 and 488~MeV~fm$^{-3}$ ($1.1$--$3.2\,\rho_0$), and the central pressures between 7.7 and 67~MeV~fm$^{-3}$ (Table~\ref{tab:props}).

	\noindent The anisotropy $\Delta\equiv p_t-p_r$ is the difference between the tangential and the radial pressure, and it has an important influence on the internal structure of compact stars. In a physically acceptable anisotropic model, $\Delta$ must be zero at the center and positive everywhere inside the star. As Fig.~\ref{fig:aniso}(a) shows, our model meets both requirements. A positive anisotropy $(p_t>p_r)$ contributes to the formation of denser and more compact stellar objects. Furthermore, the associated repulsive force generated by anisotropy counteracts gravitational attraction and thereby improves the equilibrium of the system \citep{h5}. At the surface, $\Delta$ reaches 23--49\% of the central pressure. This tangential stress has no counterpart in an isotropic fluid and should be read together with the surface layer discussed in Sec.~\ref{sec:junction}. The trace condition $\rho-p_r-2p_t>0$ \citep{p3} is not a stability requirement, and we do not impose an outward decrease of the trace. Instead, we report the normalized trace anomaly $\Delta_{\rm tr}=1/3-(p_r+2p_t)/(3\rho)$ \citep{Fujimoto2022}, which stays between $0.17$ and $0.31$ for all sources [Fig.~\ref{fig:aniso}(b)].

\par
	\noindent Following the criterion given in \citet{p4}, the equation of state (EoS) parameters, namely the ratios between pressure and density, must be smaller than one everywhere inside the star. The two ratios, $w_r=p_r/\rho$ and $w_t=p_t/\rho$, are plotted in Fig.~\ref{fig:eos}. Both remain below $0.16$, well within this bound.

\noindent The surface redshift is fixed by the exterior metric,
\begin{equation}\label{eq:zs}
z_s=C(R)^{-1/2}-1=B(R)-1 .
\end{equation}
For the configurations of Table~\ref{tab:props} we find $z_s=0.15$--$0.40$, comfortably below the limit $z_s<2$ for isotropic stars \citep{Buchdahl1959} and the limits $z_s\le3.842$  for anisotropic stars whose tangential pressure obeys the strong (dominant) energy condition \citep{Ivanov2002}.
 \par

\noindent Any realistic compact star model must satisfy the causality requirement, according to which the speed of sound cannot exceed the speed of light. For anisotropic matter distributions, the radial and tangential sound speeds are defined as
	\begin{align}
		v_i^2=\frac{dp_i}{d\rho}=\left(\frac{dp_i/dr}{d\rho/dr}\right).
	\end{align}

\noindent The no-cracking criterion of \citet{P5} requires $0\le v_r^{2}-v_t^{2}\le1$. The corresponding sound speeds and stability factor are shown in Fig.~\ref{fig:sound}. Both sound speeds satisfy $0<v_i^{2}\le1$, and $v_t^{2}<v_r^{2}$ holds throughout, so the configurations are causal and free of cracking. Only for J0740+6620 does the radial sound speed reach the conformal value $1/3$ ($v_r^{2}\le0.34$).

	\noindent Another important quantity, the adiabatic index $\Gamma_i$, is associated with the stability and stiffness of matter distribution. Following \citet{P6}, it is defined as
	\begin{align}
		\Gamma_i=\frac{\rho+p_i}{p_i}\frac{dp_i}{d\rho}.
	\end{align}
	For relativistic stars the Newtonian bound $\Gamma>4/3$ is not sufficient, so we use Eq.~(\ref{eq:gammacrit}) instead. As Fig.~\ref{fig:gamma} shows, $\Gamma_r$ and $\Gamma_t$ exceed $\Gamma_{\rm crit}$ everywhere. Even for J0740+6620 the minimum of $\Gamma_r$ is $2.42$, well above $\Gamma_{\rm crit}=1.56$.

 \par
	\noindent Behavior of the stellar matter in the interior region may also be analyzed using the standard energy conditions, which determine whether the matter distribution corresponds to ordinary or exotic matter. A physically realistic anisotropic configuration must satisfy the Null Energy Condition (NEC), Dominant Energy Condition (DEC), Weak Energy Condition (WEC) and Strong Energy Condition (SEC), given by
	\begin{itemize}
		\item NEC:\hspace{0.5cm} $\rho+p_i\ge 0$,

		\item DEC:\hspace{0.5cm}$\rho-p_i\ge 0$,

		\item WEC:\hspace{0.5cm}$\rho \ge 0$, \hspace{0.3cm} $\rho+p_i\ge 0$.

		\item SEC:\hspace{0.5cm}$\rho+p_i\ge 0$,\hspace{0.3cm}$\rho+p_r+2p_t\ge 0$,

	\end{itemize}

	\noindent Since $\rho>0$ and $p_t>0$ everywhere, $p_r\ge0$ below $r_0$, and $|p_r|\le0.0085\,p_c$ in the thin layer $r_0<r\le R$, the null, weak and strong energy conditions are satisfied throughout the star. The dominant energy condition follows from $w_r,w_t\le0.16$ (Fig.~\ref{fig:eos}).

\begin{figure}[hbt!]
	\centering
	\includegraphics[width=0.45\textwidth]{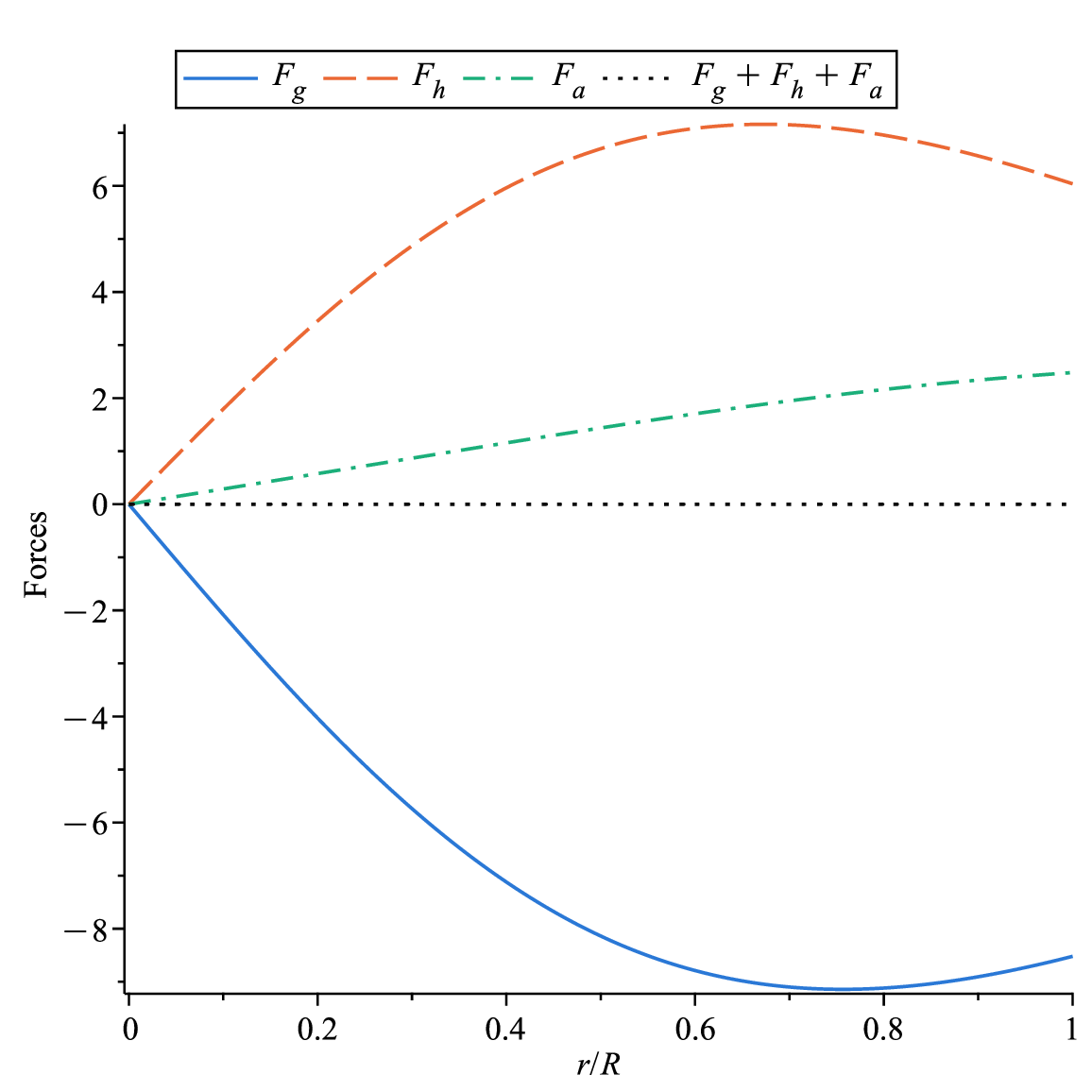}
	\caption{Force densities $F_g$, $F_h$ and $F_a$ of Eq.~(\ref{30e}) and their sum for PSR~J0740+6620 ($\alpha=0.1$~km$^{2}$, $\beta=0.1$).}
	\label{fig:forces}
\end{figure}

	\noindent Equilibrium of the stellar configuration  is governed by the Tolman-Oppenheimer-Volkoff (TOV) equation \citep{h9,P8}, expressed in Eq.~\eqref{30e} as
	\begin{align}
		\frac{2}{r}(p_t-p_r)-\frac{dp_r}{dr}-\frac{A'(r)(p_r+\rho)}{A(r)}=0. \label{30e}
	\end{align}
	Three distinct terms within the TOV equation denote the anisotropic ($F_{a}$), hydrostatic ($F_{h}$), and gravitational ($F_{g}$) force distributions. In $\mathcal{F}(\mathcal{R})= \mathcal{R} + \alpha\mathcal{R}^2$ with ordinary matter, the divergence of the field equations vanishes identically as in GR \citep{ko}. The equilibrium condition is therefore just as given in Eq.~\eqref{30e}. All the $\mathcal{F}(\mathcal{R})$ effects are already inside $F_a$, $F_h$ and $F_g$, because those were computed from the $\mathcal{F}(\mathcal{R})$ field equations. There is no separate force due to modification.
	Figure~\ref{fig:forces} shows the gravitational, hydrostatic and anisotropic force densities of Eq.~(\ref{30e}) for PSR~J0740+6620. The three contributions cancel, as they must, because for $\rho$, $p_r$ and $p_t$ obtained from the field equations Eq.~(\ref{30e}) is satisfied identically. What the figure checks is therefore our numerical implementation, not the model.
\noindent What kind of matter does the interior describe? The radial pressure goes to zero at a finite energy density $\rho_s$, and not at the low density of a crust, which is the typical behavior of self-bound matter such as deconfined quark matter \citep{Witten1984}. For the four primary pulsars of Table~\ref{tab:obs}, the linear relation $p_r\simeq v_{\rm eff}^{2}(\rho-\rho_s)$, with $v_{\rm eff}^{2}=0.22$--$0.34$ and $\rho_s=(1.05$--$1.90)\,\rho_0$, reproduces the radial equation of state to better than $0.5\%$ of the central pressure. The trace anomaly stays far from its conformal limit [Fig.~\ref{fig:aniso}(b)]. The radial sound speed squared exceeds $1/3$ only for J0740+6620, the heaviest star in the sample, where it reaches 0.34. Two-solar-mass stars require such a stiffening of the equation of state \citep{BedaqueSteiner2015}. For this reason we interpret these objects as self-bound compact stars rather than neutron stars with a crust. It has recently been argued that the small radius of PSR~J0614$-$3329 favors precisely this class of stars \citep{Shirke2025}. In Sec.~\ref{sec:obs-rhos} the surface density $\rho_s$ becomes the parameter shared by all pulsars.

\section{Comparison with NICER and Radio Observations}\label{sec:obs}

NICER measures the masses and radii of millisecond pulsars by modeling the X-ray pulses from hot spots on the stellar surface, while radio timing provides precise masses of pulsars in binary systems. These are at present the most direct radius measurements. In Secs.~\ref{sec:obs-single} and \ref{sec:obs-bounds} we check that the model is physically acceptable for each observed pulsar and use this to find the allowed ranges of $\alpha$ and $\ell$. In Sec.~\ref{sec:obs-rhos} we ask if a single physical parameter, the same for every pulsar, can account for all of them. Once this parameter is fixed, the model gives definite predictions for the mass--radius relation and the maximum mass, and these are tested against the data in Sec.~\ref{sec:obs-mr}. How much the results depend on the exterior geometry is examined in Sec.~\ref{sec:obs-exterior}. Unless stated otherwise, we use $\alpha=0.1$~km$^{2}$ and $\beta=0.1$.

\subsection{Observational data}\label{sec:obs-data}
The five pulsars with NICER radii are listed in Table~\ref{tab:obs}. For PSR~J0740+6620, J0437$-$4715 and J0614$-$3329 the masses come from radio timing and enter the X-ray analyses as priors, whereas PSR~J0030+0451 has no timing mass. Independent analyses of J0740+6620 \citep{Dittmann2024} and J0614$-$3329 \citep{Miller2026} give $12.92^{+2.09}_{-1.13}$~km and 9.88--12.77~km, respectively, in both cases consistent with the values we adopt; for an earlier analysis of J0030+0451, see \citet{Vinciguerra2024}. The radius of PSR~J1231$-$1411 is sensitive to the assumed radius prior \citep{Salmi2024J1231}, which is why we use this source only as a cross-check. Two other observations also constrain the model. The heaviest known pulsars, J0740+6620 ($2.08\pm0.07\,M_\odot$, \citet{Fonseca2021}) and J0952$-$0607 ($2.35\pm0.11\,M_\odot$, \citet{Romani2025}; first measured by \citet{Romani2022}), require a maximum mass of at least about $2.1$--$2.35\,M_\odot$, and the gravitational-wave event GW170817 gives $R_{1.4}=11.9\pm1.4$~km for a $1.4\,M_\odot$ star at 90\% credibility \citep{Abbott2018}. \citet{c51} compared a modified gravity model with J0740+6620, whereas here we use all available radius measurements. We also note that the X-ray analyses compute the light paths in a Schwarzschild exterior. For $\beta\le0.1$ the Hayward correction to $C(R)$ is $4\beta^{2}u^{4}\le1.5\times10^{-4}$ [Eq.~(\ref{23e})], which is much smaller than the systematic uncertainties of these analyses, and the published values can therefore be used directly.

\subsection{Each pulsar separately}\label{sec:obs-single}
Since the junction conditions give $a$, $b$ and $c$ in closed form from the mass, the radius and $\beta$ [Eqs.~(\ref{eq:ac}) and (\ref{eq:b}); Table~\ref{tab:consts}], a model star can be constructed for any measured mass and radius. According to Section~\ref{4s}, the stars built at the central values of all five pulsars are physically acceptable (Table~\ref{tab:props}). Because the central values are only estimates, we also ran the tests of Sec.~\ref{4s} at the four corners $(M\pm\sigma_M,\,R\pm\sigma_R)$ of the 68\% region of each pulsar. These corners cover the range $0.12\le M/R\le0.27$ of compactness. Every test is satisfied at every corner, with a surface radial stress that never exceeds $0.9\%$ of the central pressure, so the model is admissible over the full range of observed masses and radii.

\noindent On the other hand, the three constants $a$, $b$ and $c$ can always be adjusted to match a given $(M,R)$, which means that these fits cannot confirm or rule out the model. What they do provide are limits on $\alpha$ and $\ell$ (Sec.~\ref{sec:obs-bounds}). To test the model itself we need a quantity that should be the same in all pulsars (Sec.~\ref{sec:obs-rhos}).

\subsection{Allowed values of \texorpdfstring{$\alpha$ and $\ell$}{alpha and l}}\label{sec:obs-bounds}
\begin{figure}[hbt!]
\centering
\includegraphics[width=0.6\textwidth]{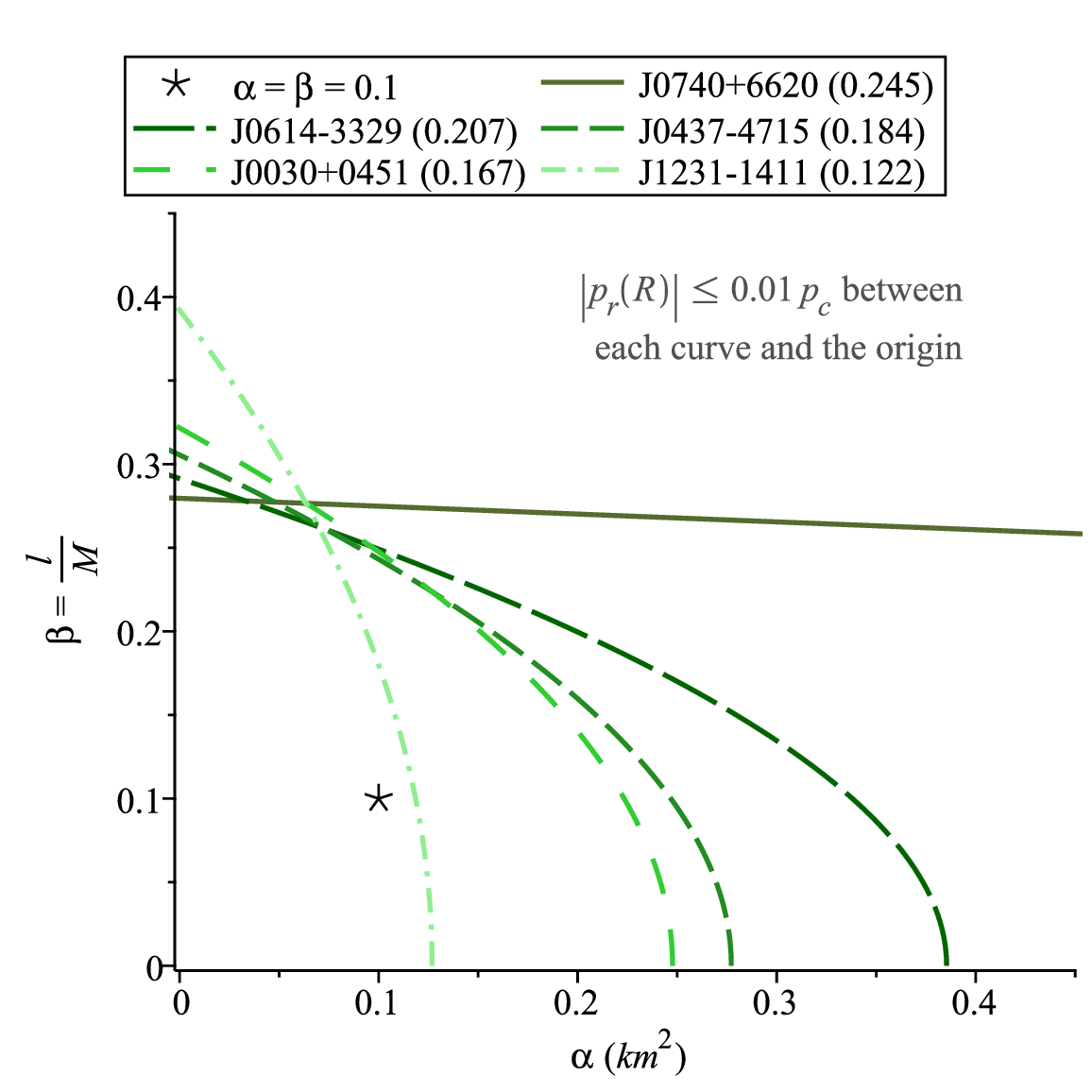}
\caption{Allowed values of $\alpha$ and $\beta=\ell/M$. For each pulsar, the radial stress at the surface satisfies $|p_r(R)|\le0.01\,p_c$ in the region between its curve and the origin. The star shows the values $\alpha=\beta=0.1$ used in this work. The bulk tests of Sec.~\ref{4s} alone would allow $\alpha$ up to 2--3~km$^{2}$, which is outside the plotted range.}
\label{fig:bounds}
\end{figure}

Two conditions restrict the parameters. One is just the requirement that the interior pass the tests of Sec.~\ref{4s}. At $\beta=0.1$ this fails for J0437$-$4715, J0030+0451, J0614$-$3329 and J1231$-$1411 above $\alpha\simeq2.9$, $3.0$, $3.4$ and $2.3$~km$^{2}$, respectively, where the tangential pressure begins to increase outward near the center ($v_t^{2}<0$); J0740+6620 gives no bound below $50$~km$^{2}$. The second condition, that the radial stress at the surface is small, is much more restrictive. Because $[K^{\mu}{}_{\mu}]=0$ fixes $b$, $p_r$ does not vanish exactly at the surface but takes a slightly negative value there [Eq.~(\ref{eq:prR})], and its magnitude grows with $\alpha$ and with $\ell^{2}$. Demanding $|p_r(R)|\le0.01\,p_c$ gives $\alpha\le0.22$, $0.25$ and $0.34$~km$^{2}$ for J0030+0451, J0437$-$4715 and J0614$-$3329 ($3.5$~km$^{2}$ for J0740+6620 and $0.12$~km$^{2}$ for J1231$-$1411) and, for $\alpha\to0$, $\beta\le0.28$--$0.32$ for the four pulsars with the best-determined radii (Fig.~\ref{fig:bounds}). We therefore adopt $\alpha\lesssim0.2$~km$^{2}$ and $\ell\lesssim0.3M$. The values $\alpha=\beta=0.1$ used in our calculations respect these bounds for all five pulsars. The 1\% tolerance is our own choice. With a tolerance of 5\%, the bound from J0030+0451 becomes $\alpha\le1.1$~km$^{2}$.

\subsection{One surface density for all pulsars}\label{sec:obs-rhos}
As noted at the end of Sec.~\ref{4s}, the radial pressure vanishes where the energy density is still finite (Fig.~\ref{fig:matter}), as it does in self-bound stars made of strange quark matter. In such stars the surface density is set by the properties of the matter itself ($\rho_s=4B$ in the MIT bag model with bag constant $B$), so it is the same for every star. This suggests using the surface energy density
\begin{equation}\label{eq:rhos}
\rho_s\equiv\rho(r=R)
\end{equation}
as a parameter common to all the pulsars. Given the measured $M$ and $R$ of a pulsar, Eqs.~(\ref{eq:ac}), (\ref{eq:b}) and (\ref{21e}) determine its $\rho_s$. To propagate the errors, we calculate $\rho_s$ again with $R$ at both ends of its 68\% interval and $M$ fixed at its central value, and with $M$ at both ends of its interval and $R$ at its central value, treating $M$ and $R$ as independent. The upper error is the quadrature sum of the shifts from the smaller $R$ and the larger $M$, both of which raise $\rho_s$, and the lower error is formed in the same way from the other two shifts. The results are listed in Table~\ref{tab:obs} in units of the nuclear saturation density, $\rho_0=2.7\times10^{14}$~g~cm$^{-3}$.

We combine the four pulsars whose radii do not depend on the assumed radius prior; PSR~J1231$-$1411, whose radius does (Sec.~\ref{sec:obs-data}), is set aside as a cross-check. The four values of $\rho_s$ range from $1.05\,\rho_0$ to $1.90\,\rho_0$. In the weighted mean we symmetrize the 68\% intervals, $\sigma_i=(\sigma_i^{+}+\sigma_i^{-})/2$, use the weights $1/\sigma_i^{2}$, and take the same $\sigma_i$ for $\chi^{2}$. This gives
\begin{equation}\label{eq:rhosjoint}
\rho_s=(1.37\pm0.16)\,\rho_0=(3.7\pm0.4)\times10^{14}~\mathrm{g\,cm^{-3}},
\end{equation}
with $\chi^{2}=2.5$ for $3$ degrees of freedom. With $\chi^{2}$ per degree of freedom of about $0.8$, the four values agree within their errors, and one surface density is enough for all of them. For PSR~J1231$-$1411 we find a lower value, $(0.82\pm0.06)\,\rho_0$, $3.2\sigma$ below Eq.~(\ref{eq:rhosjoint}). The error is small because the radius uncertainty is only $\pm0.3$~km, and if this prior-dependent value were included, it would carry about seven times the combined weight of the other four pulsars. We return to this pulsar in Sec.~\ref{sec:obs-mr}.

\subsection{Mass--radius relation and maximum mass}\label{sec:obs-mr}
\begin{figure}[hbt!]
	\begin{minipage}{0.48\textwidth}
		\includegraphics[width=\linewidth]{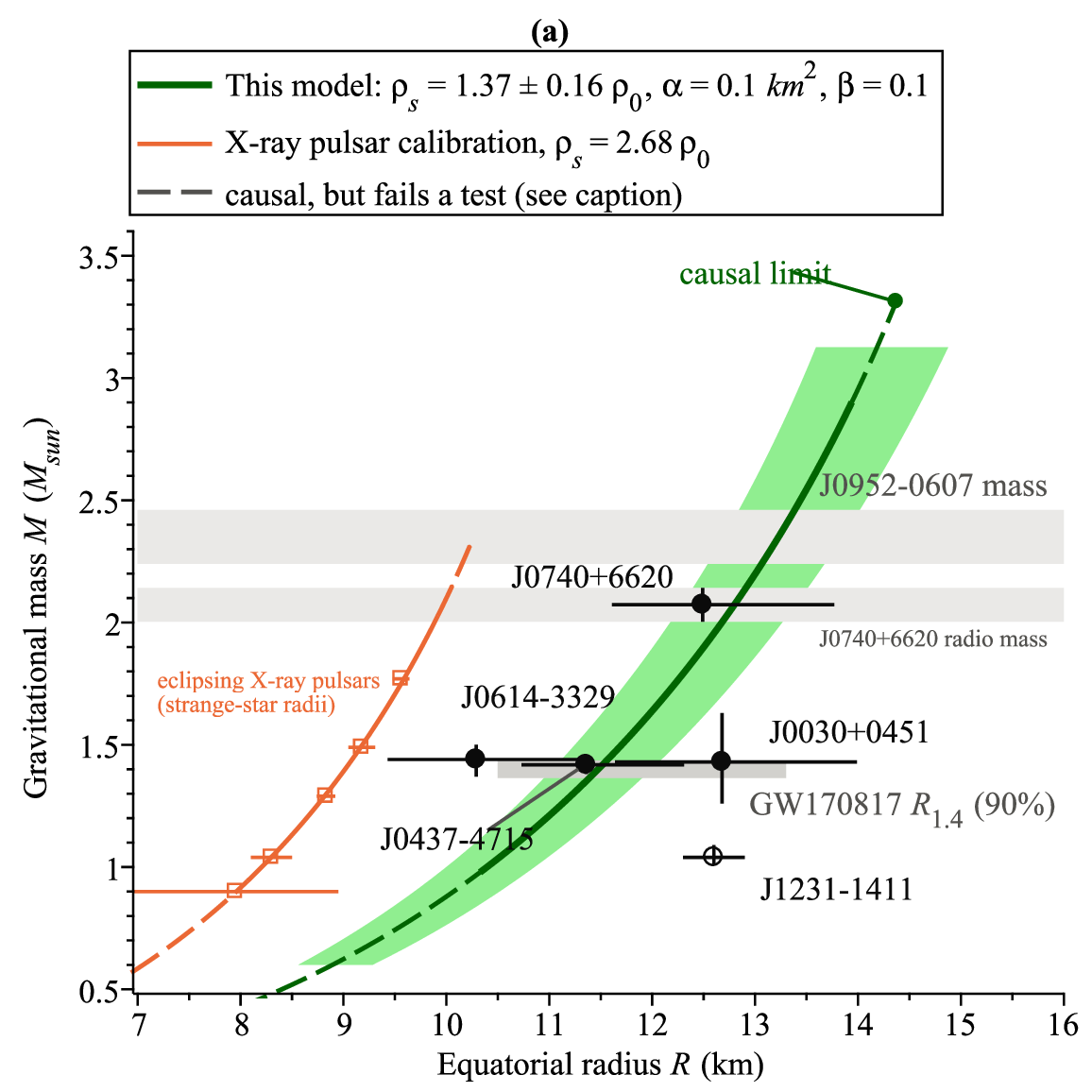}
	\end{minipage}
	\hfill
	\begin{minipage}{0.48\textwidth}
		\includegraphics[width=\linewidth]{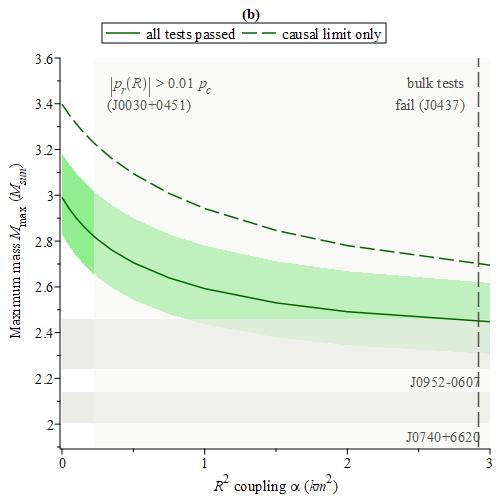}
	\end{minipage}
	\caption{(a) Mass--radius relation of the model for a fixed surface energy density $\rho_s=(1.37\pm0.16)\,\rho_0$ (green line and band; $\alpha=0.1$~km$^{2}$, $\beta=0.1$). Solid: configurations that pass all tests of Sec.~\ref{4s}. Dashed: causal configurations that fail the no-cracking test (high mass) or the surface-stress test of Sec.~\ref{sec:obs-bounds} (low mass); the dot marks the causal limit. Orange: calibration with the masses of eclipsing X-ray pulsars \citep{Rawls2011} and their strange-star radii \citep{Gangopadhyay2013} (open squares; $\rho_s=2.68\,\rho_0$). Black: NICER measurements (68\%; the J1231$-$1411 radius depends on its prior). Gray bands: radio masses of J0740+6620 and J0952$-$0607. Gray bar: GW170817 radius of a $1.4\,M_\odot$ star (90\%). (b) Maximum mass against $\alpha$ for the same range of $\rho_s$; solid: all tests passed, dashed: causal limit. In the shaded region $\alpha$ exceeds the surface-stress bound of J0030+0451; the dashed vertical line marks where the bulk tests first fail (J0437$-$4715).}
	\label{fig:mr}
\end{figure}

Once $\rho_s$ is fixed, each mass corresponds to a single radius, and the model predicts a mass--radius relation (Fig.~\ref{fig:mr}a). In GR ($\alpha=\beta=0$) the surface density is the only scale in the problem, so all masses and radii scale as $(\pi\rho_s)^{-1/2}$, much as the maximum mass of strange stars scales as $B^{-1/2}$. The maximum mass is then
\begin{equation}\label{eq:Mmax}
M_{\max}=\frac{k}{\sqrt{\pi\rho_s}},\qquad k=0.1294\ \text{(no cracking)},\qquad k=0.1470\ \text{(causality)}.
\end{equation}
The first value of $k$ corresponds to the last configuration that satisfies the no-cracking condition and the second to the last causal one; in both cases the condition first fails at the center of the star. For $\alpha,\beta\neq0$ we solve Eqs.~(\ref{eq:ac}), (\ref{eq:b}) and (\ref{eq:rhos}) numerically. At $\alpha=0.1$~km$^{2}$ the maximum mass comes out about $3\%$ below Eq.~(\ref{eq:Mmax}).\\
With the surface density of Eq.~(\ref{eq:rhosjoint}) we obtain
\begin{equation}\label{eq:preds}
R(1.04\,M_\odot)=10.5~\mathrm{km},\qquad R(1.4\,M_\odot)=11.5~\mathrm{km},\qquad R(2.08\,M_\odot)=12.8~\mathrm{km},
\end{equation}
together with a maximum mass of $M_{\max}=2.90^{+0.20}_{-0.16}\,M_\odot$ for the last configuration that passes all tests of Sec.~\ref{4s}, or $3.31^{+0.22}_{-0.19}\,M_\odot$ at the causal limit (the errors reflect those of $\rho_s$). Table~\ref{tab:compare} compares the predicted radius at the mass of each source with the measured one. The predictions agree with J0740+6620, J0437$-$4715, J0030+0451, J0614$-$3329 and the GW170817 radius to within $1.2\sigma$, and the maximum mass lies comfortably above the masses of J0740+6620 and J0952$-$0607.

\begin{table}[hbt!]
\caption{Model radius at the measured mass of each source for $\rho_s=(1.37\pm0.16)\,\rho_0$ ($\alpha=0.1$~km$^{2}$, $\beta=0.1$), compared with the measured radius (68\% intervals; 90\% for GW170817). The last column gives the difference in units of the combined uncertainty of model and measurement.}
\label{tab:compare}
\centering
\begin{tabular}{p{2.5cm}p{1.5cm}p{3cm}p{2.5cm}c}%
\toprule
Source & $M/M_\odot$ & $R_{\rm model}$ (km) & $R_{\rm obs}$ (km) & Difference\\
\midrule
J0740+6620   & 2.073 & $12.80^{+0.59}_{-0.51}$ & $12.49^{+1.28}_{-0.88}$ & $0.2\sigma$\\
J0437$-$4715 & 1.418 & $11.52^{+0.52}_{-0.44}$ & $11.36^{+0.95}_{-0.63}$ & $0.1\sigma$\\
J0030+0451   & 1.43  & $11.55^{+0.52}_{-0.45}$ & $12.68^{+1.31}_{-1.04}$ & $1.0\sigma$\\
J0614$-$3329 & 1.44  & $11.57^{+0.52}_{-0.45}$ & $10.29^{+1.01}_{-0.86}$ & $1.2\sigma$\\
GW170817     & 1.4   & $11.47^{+0.52}_{-0.44}$ & $11.9\pm1.4$            & $0.4\sigma$\\
J1231$-$1411 & 1.04  & $10.52^{+0.47}_{-0.40}$ & $12.6\pm0.3$            & $3.8\sigma$\\
\bottomrule
\end{tabular}
\end{table}

\noindent PSR~J1231$-$1411 is the one exception. Its measured radius, $12.6$~km at $1.04\,M_\odot$, is larger than that of J0437$-$4715 at $1.42\,M_\odot$. As with other self-bound stars, the radius in our model increases with mass between 1 and $2\,M_\odot$, so these two measurements cannot both be fitted with one surface density. At $1.04\,M_\odot$ the model radius is $2.1$~km ($3.8\sigma$) below the observed value. Because the J1231$-$1411 radius is prior dependent \citep{Salmi2024J1231}, we do not count this tension as a failure of the model; rather, it is a prediction that future radius measurements of light pulsars can test.\\
It is instructive to compare with the calibration often used for analytic models of this kind, which combines the masses of eclipsing X-ray pulsars \citep{Rawls2011} with radii that \citet{Gangopadhyay2013} derived from those masses using a strange-star equation of state. Applied to the present model, it gives $\rho_s=2.68\,\rho_0$, $R_{1.4}=9.0$~km and $M_{\max}=2.03\,M_\odot$ (orange curve in Fig.~\ref{fig:mr}a). The maximum mass is then below the mass of J0740+6620 and the radii are smaller than the NICER values, so this calibration is excluded by current data.\\
The predicted maximum mass exceeds the upper limit $M_{\max}\lesssim2.2$--$2.3\,M_\odot$ inferred from GW170817 \citep{Margalit2017,Rezzolla2018,Ruiz2018,Shibata2019}. These limits are based on the assumption that the merger remnant collapsed to a black hole, and on relations calibrated for isotropic neutron stars in GR, so they cannot be applied directly to anisotropic self-bound stars in $\mathcal{R}+\alpha\mathcal{R}^{2}$ gravity. Moreover, our $M_{\max}$ marks the point where the admissibility tests stop being satisfied, which is not the same as the onset of a dynamical instability. The comparison still tells us something. To have $M_{\max}=2.3\,M_\odot$, the model would need $\rho_s\simeq2.1\,\rho_0$, and with this density $R_{1.4}\simeq9.8$~km, i.e.\ $2.5\sigma$ below the NICER radius of J0437$-$4715 and outside the 90\% range of GW170817. An upper mass limit near $2.3\,M_\odot$ is therefore incompatible with the NICER radii in this model, and if such a limit were firmly established it would count against the model. On the other hand, the $\approx2.6\,M_\odot$ companion in GW190814 \citep{Abbott2020GW190814} would fit naturally in the model if it turns out to be a neutron star.\\
The $\mathcal{R}^{2}$ term lowers the maximum mass (Fig.~\ref{fig:mr}b). At the surface density of Eq.~(\ref{eq:rhosjoint}), the admissible $M_{\max}$ drops from $2.99\,M_\odot$ at $\alpha=0$ to $2.90\,M_\odot$ at $\alpha=0.1$~km$^{2}$, $2.81\,M_\odot$ at $0.25$~km$^{2}$ and $2.59\,M_\odot$ at $1$~km$^{2}$. Radii at fixed mass are much less sensitive and change by less than $0.1$~km even at $\alpha=1$~km$^{2}$.

\subsection{Role of the exterior geometry}\label{sec:obs-exterior}
The Hayward metric also lets us measure how much our results depend on the assumed exterior (Sec.~\ref{sec:exterior}). According to Eq.~(\ref{eq:betashift}), for J0740+6620 at $\beta=0.1$ the Hayward length changes $c$ by a fraction $-4.4\times10^{-4}$ and the surface redshift by $-2.0\times10^{-4}$ (a fraction $-4.9\times10^{-4}$). Even at $\beta=0.3$, the largest value allowed above, both change by only about $0.4\%$. Such changes are far below the precision of current mass and radius measurements, so the NICER constraints derived above are robust; they hold for the Hayward exterior and its Schwarzschild limit alike. What constrains the Hayward length, then, is not the present surface observables, whose sensitivity to $\ell$ is suppressed by the factor $\beta^{2}(M/R)^{3}$ of Eq.~(\ref{eq:betashift}), but the junction conditions, which give $\ell\lesssim0.3M$ (Sec.~\ref{sec:obs-bounds}).

\section{Discussion}\label{sec:disc}

\subsection{Junction conditions}\label{sec:disc-jc}
Most analytic stellar models in $\mathcal{F}(\mathcal{R})$ gravity fix the last integration constant by setting $p_r(R)=0$, exactly as one would in GR. In quadratic $\mathcal{F}(\mathcal{R})$ gravity this condition is not sufficient. Any matching has to keep the trace of the extrinsic curvature continuous \citep{Deruelle2008,Senovilla2013}, and this is equivalent to $p_r(R)=0$ only if $\alpha=0$ and the exterior is Schwarzschild. Even in GR a Hayward exterior requires $p_r(R)=-\rho_H(R)$ instead of zero, because the exterior is not vacuum. With the correct condition, $p_r$ becomes zero a little below the matching surface (17--67~m for $\alpha=\beta=0.1$), and the Ricci scalar jumps there, which produces a thin shell and a gravitational double layer \citep{Senovilla2013,Reina2016}. For these parameters the radial stress at the surface is below 1\% of the central pressure and the shell carries only 0.1--0.2\% of the stellar mass, so that the comparison with observations is not affected. However, the strongest constraint on the model comes from this surface stress, which we did not expect. It limits $\alpha$ to about $0.2$~km$^{2}$, roughly ten times less than the values at which the bulk tests of Sec.~\ref{4s} first fail. Nothing in this procedure is specific to our interior, and we would recommend that models imposing only $p_r(R)=0$ be checked against the condition $[K^{\mu}{}_{\mu}]=0$.

\subsection{Role of the exterior}\label{sec:disc-ext}
As shown in Sec.~\ref{sec:obs-exterior}, the Hayward length changes surface observables by less than $0.5\%$ for $\beta\le0.3$. This follows from the $r^{-4}$ fall-off of the Hayward correction to the Schwarzschild metric: at the surface it amounts to only a fraction of order $\beta^{2}(M/R)^{3}$ of the Schwarzschild terms. A similar estimate can be made for other regular exteriors, and for the Bardeen metric \citep{c42} it gives a relative correction of order $(g/R)^{2}$, with $g$ the Bardeen length, which is again small unless $g$ is comparable to $M$. The Hayward exterior matters for the model chiefly through the junction conditions: the surface stress there grows with $\ell^{2}$, and this leads to $\ell\lesssim0.3M$. If $\ell$ had a physical origin in nonlinear electrodynamics, it would be smaller still, since already $\beta=0.1$ requires a field of about $6\times10^{16}$~G at the surface (Sec.~\ref{sec:exterior}). A self-consistent exterior of $\mathcal{R}+\alpha\mathcal{R}^{2}$ gravity would look different from both the Hayward and the Schwarzschild metric. Outside the star the Ricci scalar decays over the Compton wavelength $\sqrt{6\alpha}$, and the mass measured at infinity can differ from the mass enclosed by the surface \citep{Yazadjiev2014}. Matching to such an exterior would remove the thin shell, and we see it as the natural next step.

\subsection{Self-bound stars and the maximum mass}\label{sec:disc-sb}
Like a self-bound star, the interior has a finite energy density at the surface and a nearly linear radial equation of state (Sec.~\ref{4s}), and the NICER data fix its surface density at $\rho_s\simeq1.4\,\rho_0$. Self-bound stars have also been put forward to explain the small radius of PSR~J0614$-$3329 \citep{Shirke2025}. Future data can test two features of the model. The first is that the radius grows with mass between 1 and $2\,M_\odot$; a less prior-dependent radius for PSR~J1231$-$1411 that confirms the present value would contradict it. The second is the maximum mass, $2.9\,M_\odot$ for $\alpha=0.1$~km$^{2}$, which lies above the upper limit of about $2.3\,M_\odot$ inferred from GW170817 for neutron stars in GR (Sec.~\ref{sec:obs-mr}). This $M_{\max}$ is the mass at which the admissibility tests first fail. In particular, it should not be confused with the maximum of an $M(R)$ curve computed from a single equation of state, because at fixed $\rho_s$ each configuration has a different pressure--density relation. In Fig.~\ref{fig:mr}(a), for example, the effective sound speed increases from $v_{\rm eff}^{2}\simeq0.20$ at $1\,M_\odot$ to $0.31$ at $2\,M_\odot$. A direct comparison with the GW170817 bound would need the dynamical stability limit, which can be obtained from a study of radial oscillations.

\subsection{Limitations}\label{sec:disc-lim}
Only the quoted 68\% intervals of $M$ and $R$ enter our analysis, and the two are treated as independent, although the published posterior samples also contain information on their correlation. Moreover, $\rho_s$ is a single point of the equation of state, so the curves of Fig.~\ref{fig:mr} form a one-parameter family of exact solutions rather than a sequence for one equation of state. Without a crust, the model can only describe self-bound stars, not ordinary neutron stars. The thin shell (Sec.~\ref{sec:junction}) and the thin layer of negative radial stress below it are consequences of our matching, and they should be absent for a richer interior or a self-consistent exterior.

\section{Conclusions}\label{5s}
We have presented an exact anisotropic solution describing self-bound compact stars in $\mathcal{F}(\mathcal{R})=\mathcal{R}+\alpha\mathcal{R}^{2}$ gravity. Its interior is of Durgapal--Bannerji type, made class one by the Karmarkar condition, and it is matched to a Hayward exterior through the junction conditions of quadratic $\mathcal{F}(\mathcal{R})$ gravity; we have also confronted the model with current mass and radius measurements. The main results of this work are the following.
\begin{enumerate}
\item The constants $a$, $b$ and $c$ are given in closed form by $M$, $R$ and $\ell$, and they do not depend on $\alpha$ [Eqs.~(\ref{eq:ac})--(\ref{eq:b}) and Table~\ref{tab:consts}]. The matching requires continuity of the trace of the extrinsic curvature in place of $p_r(R)=0$, and the jump of the Ricci scalar gives a thin shell and a gravitational double layer. For $\alpha>0$ the shell has a negative surface energy density, and its mass is 0.1--0.2\% of the stellar mass.
\item Requiring the radial stress at the surface, Eq.~(\ref{eq:prR}), to stay below 1\% of the central pressure gives $\alpha\lesssim0.2$~km$^{2}$ and $\ell\lesssim0.3M$. Without this condition, the bulk admissibility tests would allow $\alpha$ up to about 2--3~km$^{2}$.
\item PSR~J0740+6620, J0437$-$4715, J0030+0451 and J0614$-$3329 are all reproduced to within $1.2\sigma$ by one surface energy density, $\rho_s=(1.37\pm0.16)\rho_0$. With $\alpha=0.1$~km$^{2}$, the model then gives $R_{1.4}=11.5$~km and $M_{\max}=2.90^{+0.20}_{-0.16}\,M_\odot$; increasing $\alpha$ lowers the maximum mass. If the radii of eclipsing X-ray pulsars, which depend on a strange-star model, are used for the calibration, one finds $R_{1.4}\approx9$~km, and this is excluded by the data.
\item The Hayward length changes surface observables only at order $\beta^{2}(M/R)^{3}$, i.e.\ by less than $0.5\%$ within the above bound, and the results are also valid for a Schwarzschild exterior.
\item The model predicts that the radius increases with mass, which at present is in tension with the prior-dependent radius of PSR~J1231$-$1411, and its maximum mass would be excluded by a firm upper limit near $2.3\,M_\odot$. Both predictions can be tested.
\end{enumerate}
In future work, the thin shell can be avoided with a richer interior ansatz or a self-consistent scalaron exterior. It would also be useful to compute the tidal deformability for a comparison with GW170817 and to repeat the analysis with the full posterior samples.

\backmatter

\bmhead{Acknowledgements}
The authors express their gratitude to Wardah Aroosh Afzal for insightful discussions.

\bmhead{Author contributions}
Furqan Nazeer wrote the manuscript under Tooba Feroze's supervision; both authors reviewed the text.

\bmhead{Data Availability}
This study is based on published mass and radius measurements. The NICER posterior samples have been made publicly available by \citet{Salmi2024J0740}, \citet{Choudhury2024}, \citet{Kini2026}, \citet{Mauviard2025} and \citet{Salmi2024J1231}, and the code that reproduces all numbers and figures in this paper is available from the corresponding author on reasonable request.

\section*{Declarations}
\textbf{Competing interests:} The authors declare no financial conflicts of interest.\\
	\textbf{Funding:} The authors state that no external financial support or grants were obtained for this study.


\begin{thebibliography}{99}

\bibitem[Astier and Pain(2012)]{c3}
Astier, P., Pain, R.: Observational evidence of the accelerated expansion of the universe. C. R. Phys. \textbf{13}(6--7), 521--538 (2012)

\bibitem[Lange and Page(2007)]{c2}
Lange, S., Page, L.: Detecting the Expansion of the Universe through Changes in the CMB Photosphere. Astrophys. J. \textbf{671}(2), 1075 (2007)

\bibitem[Bennett et~al.(2003)]{c1}
Bennett, C.L., Hill, R.S., Hinshaw, G., Nolta, M.R., Odegard, N., Page, L., \textit{et al.}: First-year Wilkinson Microwave Anisotropy Probe (WMAP) observations: foreground emission. Astrophys. J. Suppl. Ser. \textbf{148}(1), 97 (2003)

\bibitem[Eisenstein et~al.(2005)]{c5}
Eisenstein, D.J., Zehavi, I., Hogg, D.W., Scoccimarro, R., Blanton, M.R., Nichol, R.C., \textit{et al.}: Detection of the baryon acoustic peak in the large-scale correlation function of SDSS luminous red galaxies. Astrophys. J. \textbf{633}(2), 560 (2005)

\bibitem[Planck Collaboration(2020)]{Planck2018}
Planck Collaboration: Planck 2018 results. VI. Cosmological parameters. Astron. Astrophys. \textbf{641}, A6 (2020)

\bibitem[Nojiri and Odintsov(2004)]{c6}
Nojiri, S.I., Odintsov, S.D.: Gravity assisted dark energy dominance and cosmic acceleration. Phys. Lett. B \textbf{599}(3--4), 137--142 (2004)

\bibitem[Sotiriou and Faraoni(2010)]{SotiriouFaraoni2010}
Sotiriou, T.P., Faraoni, V.: $f(R)$ theories of gravity. Rev. Mod. Phys. \textbf{82}, 451--497 (2010)

\bibitem[De Felice and Tsujikawa(2010)]{DeFeliceTsujikawa2010}
De Felice, A., Tsujikawa, S.: $f(R)$ theories. Living Rev. Relativ. \textbf{13}, 3 (2010)

\bibitem[Nojiri and Odintsov(2005)]{c7}
Nojiri, S.I., Odintsov, S.D.: Modified Gauss--Bonnet theory as gravitational alternative for dark energy. Phys. Lett. B \textbf{631}(1--2), 1--6 (2005)

\bibitem[Nojiri et~al.(2008)]{c8}
Nojiri, S.I., Odintsov, S.D., Tretyakov, P.V.: From inflation to dark energy in the non-minimal modified gravity. Prog. Theor. Phys. Suppl. \textbf{172}, 81--89 (2008)

\bibitem[Harko et~al.(2011)]{c9}
Harko, T., Lobo, F.S., Nojiri, S.I., Odintsov, S.D.: $f(R,T)$ gravity. Phys. Rev. D \textbf{84}(2), 024020 (2011)

\bibitem[Jamil et~al.(2012)]{Jamil2012}
Jamil, M., Momeni, D., Raza, M., Myrzakulov, R.: Reconstruction of some cosmological models in $f(R,T)$ gravity. Eur. Phys. J. C \textbf{72}, 1999 (2012)

\bibitem[Pretel et~al.(2025)]{h4}
Pretel, J.M.Z., \textit{et al.}: Finch--Skea dark energy stars with vanishing complexity factor. Int. J. Geom. Methods Mod. Phys. (2025)

\bibitem[Afzal and Feroze(2024)]{c17}
Afzal, W.A., Feroze, T.: Exact solutions based on Segre classification in the $f(R)$ modified theory of gravity. Eur. Phys. J. C \textbf{84}(3), 267 (2024)

\bibitem[Malik(2024)]{c16}
Malik, A.: Energy bounds in $f(R,T)$ gravity with anisotropic backgrounds. Int. J. Geom. Methods Mod. Phys. \textbf{21}(14), 2450239 (2024)

\bibitem[Singh et~al.(2024)]{c15}
Singh, J.K., Balhara, H., Singh, P.: The constrained accelerating universe in $f(R,T)$ gravity. Astron. Comput. \textbf{46}, 100795 (2024)

\bibitem[Malik et~al.(2024)]{c12}
Malik, A., Arif, A., Shamir, M.F.: Charged anisotropic compact stars in Ricci-inverse gravity. Eur. Phys. J. Plus \textbf{139}(1), 67 (2024)

\bibitem[Bhar and Pretel(2023)]{c10}
Bhar, P., Pretel, J.M.: Dark energy stars and quark stars within the context of $f(Q)$ gravity. Phys. Dark Universe \textbf{42}, 101322 (2023)

\bibitem[Jamil et~al.(2014)]{Jamil2014}
Jamil, M., Rahaman, F., Myrzakulov, R., Kuhfittig, P.K.F., Ahmed, N., Mondal, U.F.: Noncommutative wormholes in $f(R)$ gravity. J. Korean Phys. Soc. \textbf{65}, 917--925 (2014)

\bibitem[Lattimer and Prakash(2004)]{LattimerPrakash2004}
Lattimer, J.M., Prakash, M.: The physics of neutron stars. Science \textbf{304}, 536 (2004)

\bibitem[Delgaty and Lake(1998)]{c49}
Delgaty, M.S.R., Lake, K.: Physical acceptability of isolated, static, spherically symmetric, perfect fluid solutions of Einstein's equations. Comput. Phys. Commun. \textbf{115}(2--3), 395--415 (1998)

\bibitem[Durgapal and Bannerji(1983)]{dur}
Durgapal, M.C., Bannerji, R.: New analytical stellar model in general relativity. Phys. Rev. D \textbf{27}(2), 328 (1983)

\bibitem[Maurya et~al.(2026)]{mau}
Maurya, S.K., Errehymy, A., Singh, K.N., Donmez, O., Rayimbaev, J., Otaboyev, S., Turaev, Y.: Constraining $f(R,L_m,T)$ gravity with millisecond pulsar observations. Nucl. Phys. B, 117369 (2026)

\bibitem[Ditta et~al.(2023)]{dit}
Ditta, A., Tiecheng, X., Errehymy, A., Mustafa, G., Maurya, S.K.: Anisotropic charged stellar models with modified Van der Waals EoS in $f(Q)$ gravity. Eur. Phys. J. C \textbf{83}(3), 254 (2023)

\bibitem[Veneroni et~al.(2021)]{ven}
Veneroni, L.S.M., Braz, A., da Silva, M.F.A.: Compact object with a local dark energy shell. Int. J. Mod. Phys. D \textbf{30}(06), 2150039 (2021)

\bibitem[Karmarkar(1948)]{c34}
Karmarkar, K.: Gravitational metrics of spherical symmetry and class one. Proc. Indian Acad. Sci. A \textbf{27}, 56--60 (1948)

\bibitem[Ahmad and Ilyas(2025)]{ahm}
Ahmad, D., Ilyas, M.: Stellar structure and stability in $f(R)$ gravity: An analysis with the Karmarkar condition. Ann. Phys. \textbf{478}, 170032 (2025)

\bibitem[Naz et~al.(2025)]{nazt}
Naz, T., Malik, A., Akbar, K., Gul, M.Z.: Analysis of singularity-free cores: Exact two-fluid solutions for anisotropic stars in $f(R)$ gravity. Phys. Dark Universe, 102166 (2025)

\bibitem[Nazar and Abbas(2022)]{naz}
Nazar, H., Abbas, G.: Study of gravitational collapse for anisotropic Karmarkar star in minimally coupled $f(R)$ gravity. Chin. J. Phys. \textbf{79}, 124--140 (2022)

\bibitem[Olmo et~al.(2020)]{Olmo2020}
Olmo, G.J., Rubiera-Garcia, D., Wojnar, A.: Stellar structure models in modified theories of gravity: lessons and challenges. Phys. Rep. \textbf{876}, 1--75 (2020)

\bibitem[Bardeen(1968)]{c42}
Bardeen, J.: Non-singular general relativistic gravitational collapse. In: Proceedings of the 5th International Conference on Gravitation and the Theory of Relativity, p.~87 (1968)

\bibitem[Lan et~al.(2023)]{h12}
Lan, C., Yang, H., Guo, Y., Miao, Y.-G.: Regular black holes: A short topic review. Int. J. Theor. Phys. \textbf{62}, 202 (2023)

\bibitem[Pradhan and Sahoo(2024)]{h13}
Pradhan, S., Sahoo, P.K.: A comprehensive study of massive compact star admitting conformal motion under Bardeen geometry. Nucl. Phys. B \textbf{1001}, 116528 (2024)

\bibitem[Nazeer and Feroze(2025)]{h14}
Nazeer, F., Feroze, T.: Karmarkar Bardeen anisotropic model for compact stars in $\mathcal{F}(\mathcal{R},\mathcal{T})$ gravity. Astrophys. Space Sci. \textbf{370}, 33 (2025)

\bibitem[Sharif and Ibrar(2026)]{h15}
Sharif, M., Ibrar, I.: Analysis of charged compact stars with Bardeen black hole in $f(\mathfrak{Q},\mathcal{T})$ gravity. Int. J. Geom. Methods Mod. Phys. (2026), arXiv:2604.03332

\bibitem[Malik et~al.(2023)]{c48}
Malik, A., Asghar, Z., Shamir, M.F.: Krori--Barua Bardeen compact stars in $f(R,T)$ gravity. New Astron. \textbf{104}, 102071 (2023)

\bibitem[Deruelle et~al.(2008)]{Deruelle2008}
Deruelle, N., Sasaki, M., Sendouda, Y.: Junction conditions in $f(R)$ theories of gravity. Prog. Theor. Phys. \textbf{119}, 237--251 (2008)

\bibitem[Senovilla(2013)]{Senovilla2013}
Senovilla, J.M.M.: Junction conditions for $F(R)$ gravity and their consequences. Phys. Rev. D \textbf{88}, 064015 (2013)

\bibitem[Gangopadhyay et~al.(2013)]{Gangopadhyay2013}
Gangopadhyay, T., Ray, S., Li, X.-D., Dey, J., Dey, M.: Strange star equation of state fits the refined mass measurement of 12 pulsars and predicts their radii. Mon. Not. R. Astron. Soc. \textbf{431}, 3216 (2013)

\bibitem[Salmi et~al.(2024a)]{Salmi2024J0740}
Salmi, T., \textit{et al.}: The radius of the high-mass pulsar PSR J0740+6620 with 3.6 yr of NICER data. Astrophys. J. \textbf{974}, 294 (2024)

\bibitem[Choudhury et~al.(2024)]{Choudhury2024}
Choudhury, D., \textit{et al.}: A NICER view of the nearest and brightest millisecond pulsar: PSR J0437$-$4715. Astrophys. J. Lett. \textbf{971}, L20 (2024)

\bibitem[Becerra et~al.(2024)]{h10}
Becerra, L.M., Becerra-Vergara, E.A., Lora-Clavijo, F.D.: Realistic anisotropic neutron stars: Pressure effects. Phys. Rev. D \textbf{109}, 043025 (2024)

\bibitem[Starobinsky(1980)]{SM}
Starobinsky, A.A.: A new type of isotropic cosmological models without singularity. Phys. Lett. B \textbf{91}(1), 99--102 (1980)

\bibitem[N\"af and Jetzer(2010)]{NafJetzer2010}
N\"af, J., Jetzer, P.: On the $1/c$ expansion of $f(R)$ gravity. Phys. Rev. D \textbf{81}, 104003 (2010)

\bibitem[Yazadjiev et~al.(2014)]{Yazadjiev2014}
Yazadjiev, S.S., Doneva, D.D., Kokkotas, K.D., Staykov, K.V.: Non-perturbative and self-consistent models of neutron stars in R-squared gravity. J. Cosmol. Astropart. Phys. \textbf{06}, 003 (2014)

\bibitem[Singh et~al.(2017)]{new}
Singh, K.N., Pant, N., Govender, M.: Anisotropic compact stars in Karmarkar spacetime. Chin. Phys. C \textbf{41}(1), 015103 (2017)

\bibitem[Schwarzschild(1916)]{h8}
Schwarzschild, K.: On the gravitational field of a mass point according to Einstein's theory. Sitzungsber. K. Preuss. Akad. Wiss. \textbf{7}, 189--196 (1916)

\bibitem[Astashenok et~al.(2013)]{c4}
Astashenok, A.V., Capozziello, S., Odintsov, S.D.: Further stable neutron star models from $f(R)$ gravity. J. Cosmol. Astropart. Phys. \textbf{12}, 040 (2013)

\bibitem[Casado-Turri\'on(2023)]{CasadoTurrion2023}
Casado-Turri\'on, A.: Compact objects in modified gravity: junction conditions and other viability criteria. PhD thesis, Universidad Complutense de Madrid (2023). arXiv:2312.03757

\bibitem[Hayward(2006)]{h16}
Hayward, S.A.: Formation and evaporation of nonsingular black holes. Phys. Rev. Lett. \textbf{96}, 031103 (2006)

\bibitem[Fan and Wang(2016)]{FanWang2016}
Fan, Z.-Y., Wang, X.: Construction of regular black holes in general relativity. Phys. Rev. D \textbf{94}, 124027 (2016)

\bibitem[Reina et~al.(2016)]{Reina2016}
Reina, B., Senovilla, J.M.M., Vera, R.: Junction conditions in quadratic gravity: thin shells and double layers. Class. Quantum Grav. \textbf{33}, 105008 (2016)

\bibitem[Koivisto(2006)]{ko}
Koivisto, T.: A note on covariant conservation of energy--momentum in modified gravities. Class. Quantum Grav. \textbf{23}(12), 4289--4296 (2006)

\bibitem[Abreu et~al.(2007)]{P5}
Abreu, H., Hern\'andez, H., N\'u\~nez, L.A.: Sound speeds, cracking and the stability of self-gravitating anisotropic compact objects. Class. Quantum Grav. \textbf{24}(18), 4631 (2007)

\bibitem[Chandrasekhar(1964)]{Chandrasekhar1964}
Chandrasekhar, S.: The dynamical instability of gaseous masses approaching the Schwarzschild limit in general relativity. Astrophys. J. \textbf{140}, 417 (1964)

\bibitem[Moustakidis(2017)]{Moustakidis2017}
Moustakidis, C.C.: The stability of relativistic stars and the role of the adiabatic index. Gen. Relativ. Gravit. \textbf{49}, 68 (2017)

\bibitem[Kini et~al.(2026)]{Kini2026}
Kini, Y., \textit{et al.}: A NICER view of PSR J0030+0451: updated constraints from six years of NICER observations. Astrophys. J. \textbf{1005}, 201 (2026). arXiv:2602.23743

\bibitem[Mauviard et~al.(2025)]{Mauviard2025}
Mauviard, L., \textit{et al.}: A NICER view of the 1.4 solar-mass edge-on pulsar PSR J0614$-$3329. Astrophys. J. \textbf{995}, 60 (2025). doi:10.3847/1538-4357/ae145d

\bibitem[Salmi et~al.(2024b)]{Salmi2024J1231}
Salmi, T., \textit{et al.}: A NICER view of PSR J1231$-$1411: a complex case. Astrophys. J. \textbf{976}, 58 (2024)

\bibitem[Rahaman et~al.(2020)]{h5}
Rahaman, M., Singh, K.N., Errehymy, A., Rahaman, F., Daoud, M.: Anisotropic Karmarkar stars in $f(R,T)$-gravity. Eur. Phys. J. C \textbf{80}(3), 272 (2020)

\bibitem[Bondi(1999)]{p3}
Bondi, H.: The gravitational redshift from static spherical bodies. Mon. Not. R. Astron. Soc. \textbf{302}(2), 337--340 (1999)

\bibitem[Fujimoto et~al.(2022)]{Fujimoto2022}
Fujimoto, Y., Fukushima, K., McLerran, L.D., Prasza{\l}owicz, M.: Trace anomaly as signature of conformality in neutron stars. Phys. Rev. Lett. \textbf{129}, 252702 (2022)

\bibitem[Zeldovich and Smorodinsky(1962)]{p4}
Zeldovich, Y.B., Smorodinsky, Y.A.: On the upper limit on the density of neutrinos, gravitons, and baryons in the universe. Sov. Phys. JETP \textbf{14}, 647 (1962)

\bibitem[Buchdahl(1959)]{Buchdahl1959}
Buchdahl, H.A.: General relativistic fluid spheres. Phys. Rev. \textbf{116}, 1027 (1959)

\bibitem[Ivanov(2002)]{Ivanov2002}
Ivanov, B.V.: Maximum bounds on the surface redshift of anisotropic stars. Phys. Rev. D \textbf{65}, 104011 (2002)

\bibitem[Heintzmann and Hillebrandt(1975)]{P6}
Heintzmann, H., Hillebrandt, W.: Neutron stars with an anisotropic equation of state: mass, redshift and stability. Astron. Astrophys. \textbf{38}, 51--55 (1975)

\bibitem[Oppenheimer and Volkoff(1939)]{h9}
Oppenheimer, J.R., Volkoff, G.M.: On massive neutron cores. Phys. Rev. \textbf{55}(4), 374 (1939)

\bibitem[Tolman(1939)]{P8}
Tolman, R.C.: Static solutions of Einstein's field equations for spheres of fluid. Phys. Rev. \textbf{55}(4), 364 (1939)

\bibitem[Witten(1984)]{Witten1984}
Witten, E.: Cosmic separation of phases. Phys. Rev. D \textbf{30}, 272 (1984)

\bibitem[Bedaque and Steiner(2015)]{BedaqueSteiner2015}
Bedaque, P., Steiner, A.W.: Sound velocity bound and neutron stars. Phys. Rev. Lett. \textbf{114}, 031103 (2015)

\bibitem[Shirke et~al.(2025)]{Shirke2025}
Shirke, S., Maiti, R., Chatterjee, D.: PSR J0614$-$3329: a NICER case for strange quark stars. arXiv:2508.02652 (2025)

\bibitem[Dittmann et~al.(2024)]{Dittmann2024}
Dittmann, A.J., \textit{et al.}: A more precise measurement of the radius of PSR J0740+6620 using updated NICER data. Astrophys. J. \textbf{974}, 295 (2024)

\bibitem[Miller et~al.(2026)]{Miller2026}
Miller, M.C., Dittmann, A.J., Holt, I.M., \textit{et al.}: The radius of the neutron star PSR J0614$-$3329 from NICER data. arXiv:2609.00965 (2026)

\bibitem[Vinciguerra et~al.(2024)]{Vinciguerra2024}
Vinciguerra, S., \textit{et al.}: An updated mass--radius analysis of the 2017--2018 NICER data set of PSR J0030+0451. Astrophys. J. \textbf{961}, 62 (2024)

\bibitem[Fonseca et~al.(2021)]{Fonseca2021}
Fonseca, E., \textit{et al.}: Refined mass and geometric measurements of the high-mass PSR J0740+6620. Astrophys. J. Lett. \textbf{915}, L12 (2021)

\bibitem[Romani et~al.(2025)]{Romani2025}
Romani, R.W., Beleznay, M., Filippenko, A.V., Brink, T.G., Zheng, W.: PSR J0952$-$0607: tightening a record-high neutron star mass. Astrophys. J. \textbf{996}, 101 (2026). arXiv:2512.05099

\bibitem[Romani et~al.(2022)]{Romani2022}
Romani, R.W., Kandel, D., Filippenko, A.V., Brink, T.G., Zheng, W.: PSR J0952$-$0607: the fastest and heaviest known Galactic neutron star. Astrophys. J. Lett. \textbf{934}, L17 (2022)

\bibitem[Abbott et~al.(2018)]{Abbott2018}
Abbott, B.P., \textit{et al.} (LIGO Scientific and Virgo Collaborations): GW170817: Measurements of neutron star radii and equation of state. Phys. Rev. Lett. \textbf{121}, 161101 (2018)

\bibitem[Nashed(2023)]{c51}
Nashed, G.G.L.: Confront $f(R,T)=R+\beta T$ modified gravity with the massive pulsar PSR J0740+6620. Eur. Phys. J. C \textbf{83}(8), 698 (2023)

\bibitem[Rawls et~al.(2011)]{Rawls2011}
Rawls, M.L., \textit{et al.}: Refined neutron-star mass determinations for six eclipsing X-ray pulsar binaries. Astrophys. J. \textbf{730}, 25 (2011)

\bibitem[Margalit and Metzger(2017)]{Margalit2017}
Margalit, B., Metzger, B.D.: Constraining the maximum mass of neutron stars from multi-messenger observations of GW170817. Astrophys. J. Lett. \textbf{850}, L19 (2017)

\bibitem[Rezzolla et~al.(2018)]{Rezzolla2018}
Rezzolla, L., Most, E.R., Weih, L.R.: Using gravitational-wave observations and quasi-universal relations to constrain the maximum mass of neutron stars. Astrophys. J. Lett. \textbf{852}, L25 (2018)

\bibitem[Ruiz et~al.(2018)]{Ruiz2018}
Ruiz, M., Shapiro, S.L., Tsokaros, A.: GW170817, general relativistic magnetohydrodynamic simulations, and the neutron star maximum mass. Phys. Rev. D \textbf{97}, 021501 (2018)

\bibitem[Shibata et~al.(2019)]{Shibata2019}
Shibata, M., Zhou, E., Kiuchi, K., Fujibayashi, S.: Constraint on the maximum mass of neutron stars using GW170817 event. Phys. Rev. D \textbf{100}, 023015 (2019)

\bibitem[Abbott et~al.(2020)]{Abbott2020GW190814}
Abbott, R., \textit{et al.} (LIGO Scientific and Virgo Collaborations): GW190814: Gravitational waves from the coalescence of a 23 solar mass black hole with a 2.6 solar mass compact object. Astrophys. J. Lett. \textbf{896}, L44 (2020)

\end{thebibliography}
\end{document}